\documentclass{aa}
\usepackage{txfonts}
\usepackage{graphicx}
\usepackage{caption}
\usepackage{subcaption}
\usepackage{color}
\usepackage{tablefootnote}

\usepackage{tikz}
\usetikzlibrary{arrows,shapes,backgrounds}

\begin{document} 
\titlerunning{Galaxy properties of type 1 and 2 X-ray selected AGN and comparison among different classification criteria}
\authorrunning{Mountrichas et al. }
\titlerunning{Galaxy properties of type 1 and 2 X-ray selected AGN and comparison among different classification criteria}

\title{Galaxy properties of type 1 and 2 X-ray selected AGN and comparison among different classification criteria}

\author{G. Mountrichas\inst{1}, V. Buat\inst{2,3}, I. Georgantopoulos\inst{4}, G. Yang\inst{5,6}, V. A. Masoura\inst{4,7}, M. Boquien\inst{8}, D. Burgarella\inst{2}}
          
    \institute {Instituto de Fisica de Cantabria (CSIC-Universidad de Cantabria), Avenida de los Castros, 39005 Santander, Spain
              \email{gmountrichas@gmail.com}           
             \and
             Aix Marseille Univ, CNRS, CNES, LAM Marseille, France. 
                \email{ veronique.buat@lam.fr}  
              \and
                 Institut Universitaire de France (IUF)
                 \and
                 National Observatory of Athens, V.  Paulou  \& I.  Metaxa, 15 236 Penteli, Greece
                 \and
                 Department of Physics and Astronomy, Texas A\&M University, College Station, TX 77843-4242, USA 
               \and
                George P. and Cynthia Woods Mitchell Institute for Fundamental Physics and Astronomy, Texas A\&M University, College Station, TX 77843-4242, USA 
                \and
                Section of Astrophysics, Astronomy and Mechanics, Department of Physics, Aristotle University of Thessaloniki, 54 124, Thessaloniki, Greece
                \and
                Centro de Astronom\'ia (CITEVA), Universidad de Antofagasta, Avenida Angamos 601, Antofagasta, Chile}

\abstract {We present analyses of host galaxy properties of type 1 and type 2 X-ray selected AGNs in the XMM-XXL field, which have available optical spectroscopic classification. We model their optical to far-infrared spectral energy distributions (SEDs) using the X-CIGALE code. X-CIGALE allows the fitting of X-ray flux and accounts for the viewing angle of dusty torus and the attenuation from polar dust. By selecting matched type 1 and 2 subsamples in the X-ray luminosity and redshift parameter space, we find that both types live in galaxies with similar star formation. However, type 2 AGN tend to reside in more massive systems ($10.87^{+0.06}_{-0.12}\,\rm M_\odot$) compared to their type 1 counterparts ($10.57^{+0.20}_{-0.12}\,\rm M_\odot$). In the second part of our analysis, we compare the spectroscopic classification with that from the SED fitting. X-CIGALE successfully identifies all spectroscopic type 2 sources either by estimating an inclination angle that corresponds to edge on viewing of the source or by measuring increased polar dust in these systems. $\sim 85\%$ of spectroscopic type 1 AGN are also identified as such, based on the SED fitting analysis. There is a small number of sources ($\sim 15\%$ of the sample), that present broad lines in their spectra, but show strong indications of obscuration, based on SED analysis. These, could be systems that are viewed face on and have an extended dust component along the polar direction. The performance of X-CIGALE in classifying AGN is similar at low and high redshifts, under the condition that there is sufficient photometric coverage. Finally, usage of optical/mid-IR colour criteria to identify optical red AGN (${\it{u}}-\rm W3$), suggests that these criteria are better suited for IR selected AGN and their efficiency drops for the low to moderate luminosity sources included in X-ray samples.} 

   
\maketitle   

\section{Introduction}

Active Galactic Nuclei (AGN) play an important role in galaxy evolution. They are powered by accretion onto the supermassive black hole (SMBH) located at the centre of galaxies. The AGN-galaxy co-evolution is governed by SMBH feeding mechanism(s) and AGN feedback. To decipher this interplay between the active SMBH and its host galaxy, it is important to shed light on the AGN structure. One of the most important aspects of this pursuit is to understand the physical difference between obscured and unobscured AGN. 

According to the simplest form of the unification model \citep[e.g.][]{Antonucci1993, Urry1995, Nenkova2002, Hoenig2006, Schartmann2008, Netzer2015}, AGN are surrounded by a dusty gas torus structure that absorbs radiation emitted from the nucleus, i.e. the SMBH and the accretion disc around it. This absorbed radiation is then re-emitted at larger (infrared) wavelengths. In this scenario, the viewing angle that the AGN is observed determines whether the source is observed as obscured or unobscured. When the AGN is viewed face-on then the source is classified as unobscured (type 1), while when the AGN is observed edge-on the source is characterized as obscured (type 2). Thus, the classification of AGN into different types is purely a geometrical effect. Although, more complex AGN structures have been proposed to explain e.g. the diversity of classification at different wavelengths (e.g., X-ray vs. optical classifications) under the unified scheme of AGN \citep[e.g.,][]{Ogawa2021, Arredondo2021}, the inclination angle remains the determinant factor for classifying obscured and unobscured AGN. 

However, in the context of the evolutionary models, different AGN types are attributed to SBMH and galaxies being observed at different phases. The core idea of these models is that obscured AGN are observed during an early phase, when the SMBH is still weak and incapable of expelling the surrounding gas and dust that has been pushed towards the galactic centre. This material feeds the AGN that eventually becomes powerful enough to push away the surrounding material \citep[e.g.][]{Ciotti1997, Hopkins2006, Somerville2008}. Under this scheme, obscuration is not solely related to the AGN torus, but also to absorbing content at galactic scales \citep[e.g.][]{Circosta2019, Malizia2020}.

Study of the two AGN populations can shed light on many different aspects of the AGN-galaxy interplay. Under the evolutionary scheme, study of the obscured and unobscured AGN populations has further implications regarding e.g., the relation between BH growth and the growth of the host galaxy and the lifetime of each classification type \citep[e.g.][]{Hickox2011, Ballantyne2017b}. However, we first need to understand what is the nature of obscured and unobscured AGN, i.e., whether the two AGN types are a geometrical effect or represent different stages of galaxy evolution. A popular approach to answer this question is to compare the host galaxy properties of obscured and unobscured AGN. If the two AGN populations live in similar environments this would provide support to the unification model whereas if they reside in galaxies of different properties, it would suggest that they are observed at different evolutionary phases. 


Previous works have selected AGN at different wavelengths and have applied different obscuration criteria to classify sources. Most studies that examined the host galaxy properties of X-ray selected AGN and classified them into obscured and unobscured using X-ray criteria, e.g. the value of the hydrogen column density, N$_H$, agree that both X-ray absorbed and unabsorbed AGN live in galaxies with similar stellar mass, M$_*$, and star-formation rate, SFR, \citep[][, but see Lanzuisi et al. 2017]{Merloni2014, Masoura2021, Mountrichas2021b}. \cite{Chen2015} used IR selected AGN in the Bo$\rm \ddot{o}$tes field and classified their sources using optical/mid-IR colours. Their analysis showed that type 2 AGN have higher IR star formation luminosities (i.e., higher SFR) compared to type 1 AGN, by a factor of $\sim 2$. \cite{Zou2019} used X-ray AGN and classified them based on optical spectra, morphology and optical variability. Their analysis showed that type 2 AGN reside in more massive systems than their type 1 counterparts. Although the results of \cite{Chen2015} cannot be directly compared to those from X-ray studies, since their AGN selection is different, the results of \cite{Zou2019} suggest that although host galaxy properties are similar for X-ray absorbed and unabsorbed AGN, spectroscopically obscured X-ray sources reside in more massive systems than unobscured sources.


In this work, we use $\sim 2500$ spectroscopic, X-ray selected AGN in the {\it{XMM}}-XXL field that have available classification, based on their optical spectra from the literature \citep{Menzel2016}. We apply SED fitting using the X-CIGALE code to measure the host galaxy properties of the two AGN types and compare them. In the second part of the analysis, we use the estimated inclination angle of each source from the SED fitting and compare it with the spectroscopic classification. Our goal is to examine the reliability of X-CIGALE classification and its effect on the accuracy on the measurements of the host galaxy properties. We also examine the efficiency of optical/mid-IR colour criteria and specifically the ${\it{u}}-\rm W3$ criterion of \cite{Hickox2017} in selecting obscured sources in an X-ray selected sample. 


Throughout this work, we assume a flat $\Lambda$CDM cosmology with $H_ 0=69.3$\,Km\,s$^{-1}$\,Mpc$^{-1}$ and $\Omega _ M=0.286$.

\section{Sample}
\label{sec_sample}

We use spectroscopic X-ray AGN from the XMM-XXL field \citep{Pierre2016}. XXL is a medium depth X-ray survey, with sensitivity of $\sim$  6 $\times$ $10^{-15}$ erg\,cm\,$^{-2}$\,s$^{-1}$ in the [0.5-2] keV band for point-like sources and exposure time of about 10\,ks per XMM pointing. It consists of two 25\,deg$^2$ extragalactic fields. The data used in this work come from the equatorial sub-region of the XXM-XXL North. 8,445 X-ray sources have been observed in this field. 5,294 have SDSS counterparts and spectroscopic redshifts are available for 2,512 sources. The catalogue is presented in detail in \cite{Menzel2016}. 

In our analysis, we use 2,512 spectroscopic X-ray selected AGN that lie within a redshift range of $0<z<5$ ($94\%$ are at $z<2.5$). The available photometry is described in Section 2 of \cite{Mountrichas2021}. In brief, additionally to the optical (SDSS) photometry available, the sources have also near-IR and mid-IR photometry from the Visible and Infrared Survey Telescope for Astronomy \citep[VISTA;][]{Emerson2006} and the allWISE \citep{Wright2010} datasets. We also used catalogues produced by the HELP\footnote{The {\it Herschel} Extragalactic Legacy Project (HELP; http://herschel.sussex.ac.uk/) is a European-funded project to analyse all the cosmological fields observed with the {\it Herschel} satellite. All the HELP data products can be accessed on HeDaM (http://hedam.lam.fr/HELP/).} collaboration to complement our mid-IR photometry with {{\it{Spitzer}} \citep{Werner2004} observations, and we added far-IR counterparts. Only the MIPS and SPIRE fluxes were considered, given the much lower sensitivity of the PACS observations for this field \citep{Oliver2012}. The W1 and W2 photometric bands of WISE nearly overlap with IRAC1 and IRAC2 from {\it Spitzer}. When a source had been detected by both IR surveys, we only considered the photometry with the highest signal-to-noise ratio (S/N). Similarly, when both W4 and MIPS photometry is available, we only considered the latter due to the higher sensitivity of {\it Spitzer} compared to WISE.

Robust measurements of galaxy properties are essential in our analysis. For this reason, we require sources to have the following photometry available: {\it{u, g, r, i, z}}, W1 or IRAC1, W2 or IRAC2, W3, W4 or MIPS1. The resulted sample consists of 2,134 sources. UV photometry allows tracing the young stellar population. At $\rm z>0.5$, the {\it{u}} optical band is redshifted to rest-frame wavelength $<2000\,\AA$, allowing observation of the emitted radiation from young stars. At $\rm z<0.5$, shorter wavelengths are required (e.g. GALEX). GALEX photometric data are available for only 216 out of the 2,512 X-ray sources ($\sim 8\%$). Since the vast majority of AGN does not have GALEX photometry, we choose not to include it in the construction of the SEDs. Therefore, at $\rm z<0.5$, in the absence of UV photometry, we only keep sources that have available {\it{Herschel}} photometry. This reduces the number of sources to 1,897. Finally, we exclude from our analysis sources that do not have reliable optical spectral classifications (for details see Section \ref{sec_optspectra}). This results in 1,577 X-ray selected AGN. 




\begin{table*}
\caption{The models and the values for their free parameters used by X-CIGALE for the SED fitting of our galaxy sample. } 
\centering
\setlength{\tabcolsep}{1.mm}
\begin{tabular}{cc}
       \hline
Parameter &  Model/values \\
	\hline
\multicolumn{2}{c}{Star formation history: delayed model and recent burst} \\
Age of the main population & 1500, 2000, 3000, 4000, 5000 Myr \\
e-folding time & 200, 500, 700, 1000, 2000, 3000, 4000, 5000 Myr \\ 
Age of the burst & 50 Myr \\
Burst stellar mass fraction & 0.0, 0.005, 0.01, 0.015, 0.02, 0.05, 0.10, 0.15, 0.18, 0.20 \\
\hline
\multicolumn{2}{c}{Simple Stellar population: Bruzual \& Charlot (2003)} \\
Initial Mass Function & Chabrier (2003)\\
Metallicity & 0.02 (Solar) \\
\hline
\multicolumn{2}{c}{Galactic dust extinction} \\
Dust attenuation recipe & Charlot \& Fall (2000)    \\
V-band attenuation $A_V$ & 0.2, 0.3, 0.4, 0.5, 0.6, 0.7, 0.8, 0.9, 1, 1.5, 2, 2.5, 3, 3.5, 4 \\ 
\hline
\multicolumn{2}{c}{Galactic dust emission: Dale et al. (2014)} \\
$\alpha$ slope in $dM_{dust}\propto U^{-\alpha}dU$ & 2.0 \\
\hline
\multicolumn{2}{c}{AGN module: SKIRTOR} \\
Torus optical depth at 9.7 microns $\tau _{9.7}$ & 3.0, 7.0 \\
Torus density radial parameter p ($\rho \propto r^{-p}e^{-q|cos(\theta)|}$) & 1.0 \\
Torus density angular parameter q ($\rho \propto r^{-p}e^{-q|cos(\theta)|}$) & 1.0 \\
Angle between the equatorial plan and edge of the torus & $40^{\circ}$ \\
Ratio of the maximum to minimum radii of the torus & 20 \\
Viewing angle  & $30^{\circ}\,\,\rm{(type\,\,1)},70^{\circ}\,\,\rm{(type\,\,2)}$ \\
AGN fraction & 0.0, 0.1, 0.2, 0.3, 0.4, 0.5, 0.6, 0.7, 0.8, 0.9, 0.99 \\
Extinction law of polar dust & SMC \\
$E(B-V)$ of polar dust & 0.0, 0.2, 0.4 \\
Temperature of polar dust (K) & 100 \\
Emissivity of polar dust & 1.6 \\
\hline
\multicolumn{2}{c}{X-ray module} \\
AGN photon index $\Gamma$ & 1.8 \\
Maximum deviation from the $\alpha _{ox}-L_{2500 \AA}$ relation & 0.2 \\
LMXB photon index & 1.56 \\
HMXB photon index & 2.0 \\
\hline
Total number of models & 320,760,000 \\
\hline
\label{table_cigale}
\end{tabular}
\tablefoot{For the definition of the parameters see Section \ref{sec_cigale}.}
\end{table*}

\section{Analysis}
In this Section, we describe the SED fitting analysis we follow to measure galaxy properties. We present our criteria to select only those sources with robust galaxy properties calculations that are included in our final sample. Finally, we examine the reliability of the SED fitting measurements. 

\subsection{SED analysis using X-CIGALE}
\label{sec_cigale}

We measure the host galaxy properties of X-ray AGN in our sample, by applying SED fitting using the X-CIGALE code \citep{Yang2020}. X-CIGALE is a newly developed branch of the CIGALE fitting code \citep{Boquien2019} that adds some important new features. The new algorithm has the ability to account for extinction of the ultraviolet (UV) and optical emission in the poles of AGN and models the X-ray emission of galaxies. For the latter, it requires the intrinsic X-ray fluxes, i.e., X-ray fluxes corrected for X-ray absorption. The improvements that these new features add in the fitting process are described in detail in \cite{Yang2020} and \cite{Mountrichas2021}.

For the SED fitting process, we use the intrinsic X-ray fluxes estimated in \cite{Mountrichas2021}. These measurements are based on hardness ratio estimations, calculated via a Bayesian approach called Bayesian Estimation of Hardness Ratios code \cite[BEHR;][]{Park2006}. The details are presented in Section 3.1 of \cite{Mountrichas2021}. X-CIGALE uses the $\rm \alpha _{ox}-L_{2500\AA}$ relation of \cite{Just2007} to connect the X-ray flux with the AGN emission at 2500\AA. We adopt a maximal value of $|\Delta \alpha _{ox}|_{max}=0.2$ that accounts for a $\approx 2\,\sigma$ scatter in the above relation. Photon index, $\Gamma$, i.e., the slope of the X-ray spectrum, is set to 1.8.

AGN emission is modelled using the SKIRTOR templates \citep{Stalevski2012, Stalevski2016}. SKIRTOR assumes a clumpy two-phase torus model, based on 3D radiation-transfer. The model presented in Stalevski et al. is used for the UV to far-IR AGN emission with some modifications: the accretion disc is updated with the spectral energy distribution of \cite{Feltre2012} and dust emission and extinction is added in the poles of the AGN. For more details we refer to \cite{Yang2020}. AGN fraction is defined as the ratio of the AGN IR emission to the total IR emission of the galaxy, i.e., the integrated luminosity from 8-1000\,$\mu$m in rest-frame. The extinction due to polar dust is modelled as a dust screen geometry and a grey-body dust re-emission. The amount of extinction is measured with the $E_{B-V}$ parameter. We adopt the Small Magellanic Cloud extinction curve \citep[SMC;][]{Prevot1984}. Re-emitted grey-body dust is parameterized with a temperature of 100\,K and emissivity index of 1.6. Dust temperature has likely a wide distribution, but the its effect on the SED fitting parameters is minimal \citep[][Buat et al. in prep.]{Mountrichas2021}. The galaxy component is fitted using a delayed star formation history (SFH) with the functional form SFR\,$\propto t\times \exp(-t/{\tau})$. A star formation burst is also considered and modelled as a constant ongoing star formation no longer than 50\,Myr. The burst is  superimposed to the delayed SFH \citep{Buat2019}. Stellar emission is modelled using the \cite{Bruzual_Charlot2003} single stellar populations template. The initial mass function (IMF) of \cite{Chabrier2003} is adopted. Metallicity is fixed to 0.02. Stellar emission is attenuated following the \cite{Charlot_Fall_2000} recipe. We adopt a value of $\mu =0.5$, where $\mu$ is the ratio of the total attenuation undergone by stars older than 10\,Myr to that undegone by stars younger than 10\,Myr \citep{Malek2018, Buat2019}. The IR SED of the dust heated by stars is implemented with the \cite{Dale2014} library, without the AGN component. 

\begin{figure}
\centering
\begin{subfigure}[b]{0.5\textwidth}
   \includegraphics[width=1\linewidth, height=7.2cm]{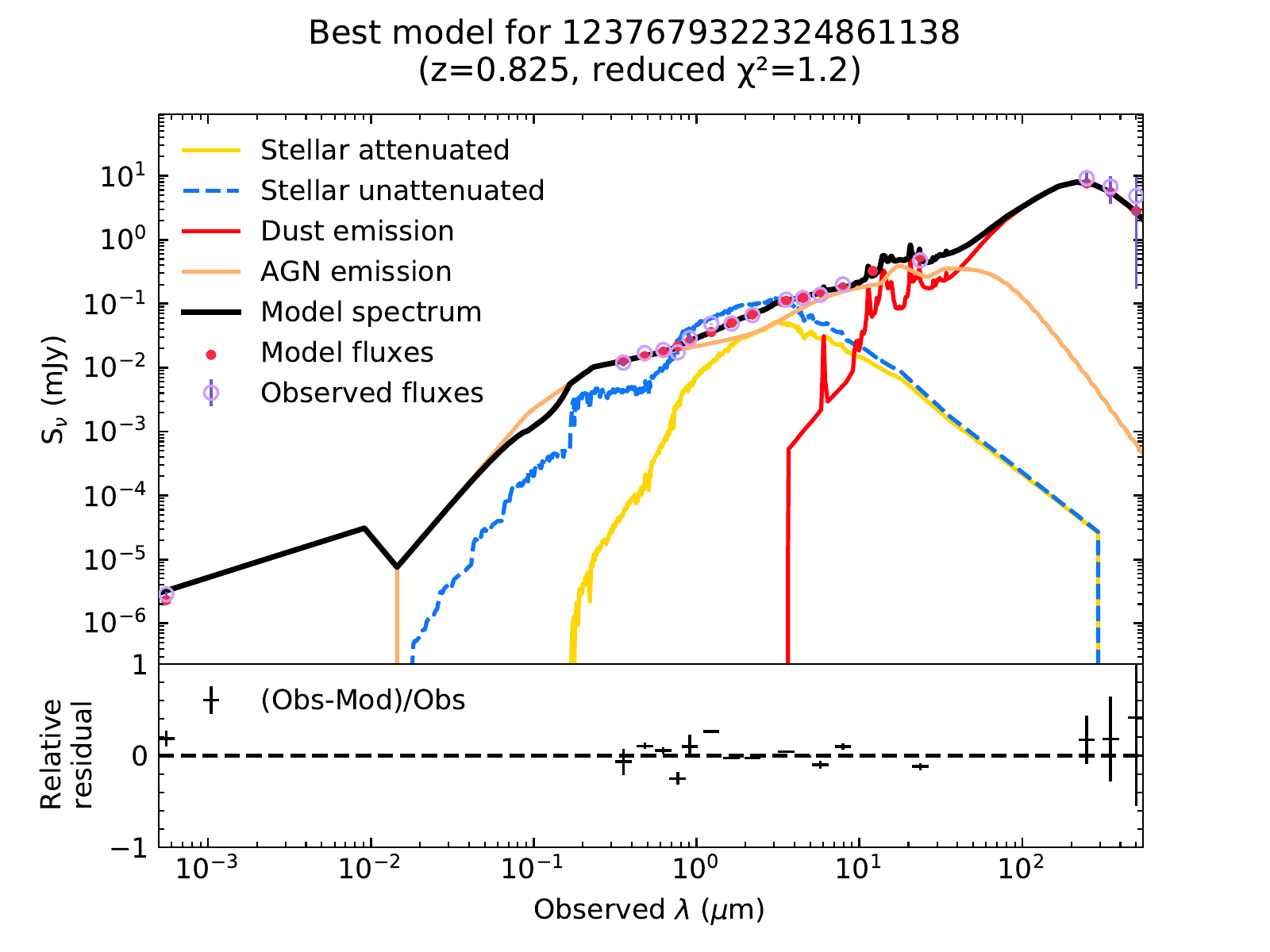}
   \label{} 
\end{subfigure}

\begin{subfigure}[b]{0.5\textwidth}
   \includegraphics[width=1\linewidth, height=7.2cm]{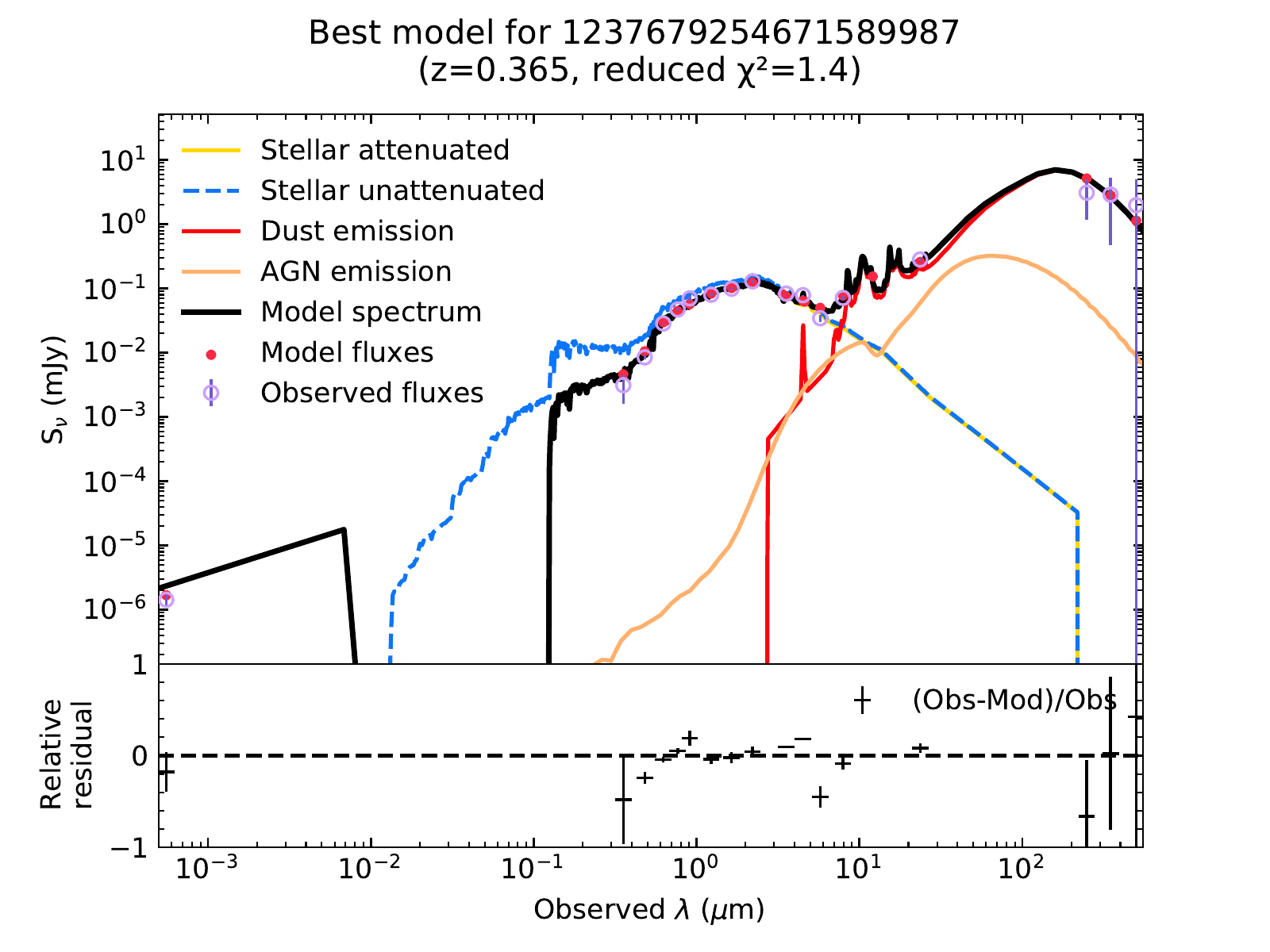}
   \label{}
\end{subfigure}

\caption{Examples of SEDs from sources that satisfy our selection criteria (see Section \ref{sec_unreliable}). A source classified as type 1 based on SED fitting is presented in the top panel. A type 2 AGN is presented in the bottom panel. (see text for more details on SED classification).}
\label{fig_SEDs_good}
\end{figure}

\subsection{Selection of galaxy SEDs with secure fits}
\label{sec_unreliable}

For each parameter estimated via SED fitting, the algorithm provides two estimates. One is calculated using the best-fit model (best value) and the other estimate of a parameter corresponds to the likelihood weighted mean value measured from  its  probability density function marginalized over all the other parameters (bayes value). The weight is based on the likelihood, exp\,(-$\chi ^2/2$), associated with each model \citep{Boquien2019}. A large difference between these two values is an indication that the probability density function (PDF) is asymmetric or multi-peaked. Therefore, to exclude from our analysis sources with unreliable SFR and M$_*$ measurements, we consider only X-ray sources with $\rm \frac{1}{5}\leq \frac{SFR_{best}}{SFR_{bayes}} \leq 5$ and $\rm \frac{1}{5}\leq \frac{M_{*, best}}{M_{*, bayes}} \leq 5$, where SFR$\rm _{best}$, M$\rm _{*, best}$ are the best fit values of SFR and M$_*$, respectively and SFR$\rm _{bayes}$ and M$\rm _{*, bayes}$ are the Bayesian values, estimated by X-CIGALE. These criteria, reduce the number of X-ray sources to 1292 (from 1577). The choice of the limits is empirical. Allowing for more strict or loose lower and upper boundaries, e.g. $0.1-0.33$ and $3-10$ changes the size of our catalogue by less than $\pm 0.1\%$ and thus does not affect our results and conclusions. Additionally, 91 sources ($\approx 7\%$ of the sample) with $\chi ^2_{red}>5$ are excluded from our analysis (see Appendix for more details). Fig. \ref{fig_SEDs_good} presents examples of SEDs from sources that satisfy the aforementioned criteria. Finally, among the 1201 (1292-91) AGN, we only consider sources with SFR and M$_*$ measurements that have statistical significance, $S/N>2$. $S/N$, is defined as $S/N=\frac{value}{error}$, where the error in the denominator is the error of the parameter, estimated by X-CIGALE. This requirement reduces further the AGN sample and is discussed in more detail in Section \ref{sec_comparison_host_results}. The combination of these selection criteria allows us to compare host galaxy properties of spectroscopic type 1 and 2 X-ray AGN, using only sources with the most robust host galaxy properties measurements.

\begin{figure}
\centering
  \centering
  \includegraphics[width=1.\linewidth, height=7.2cm]{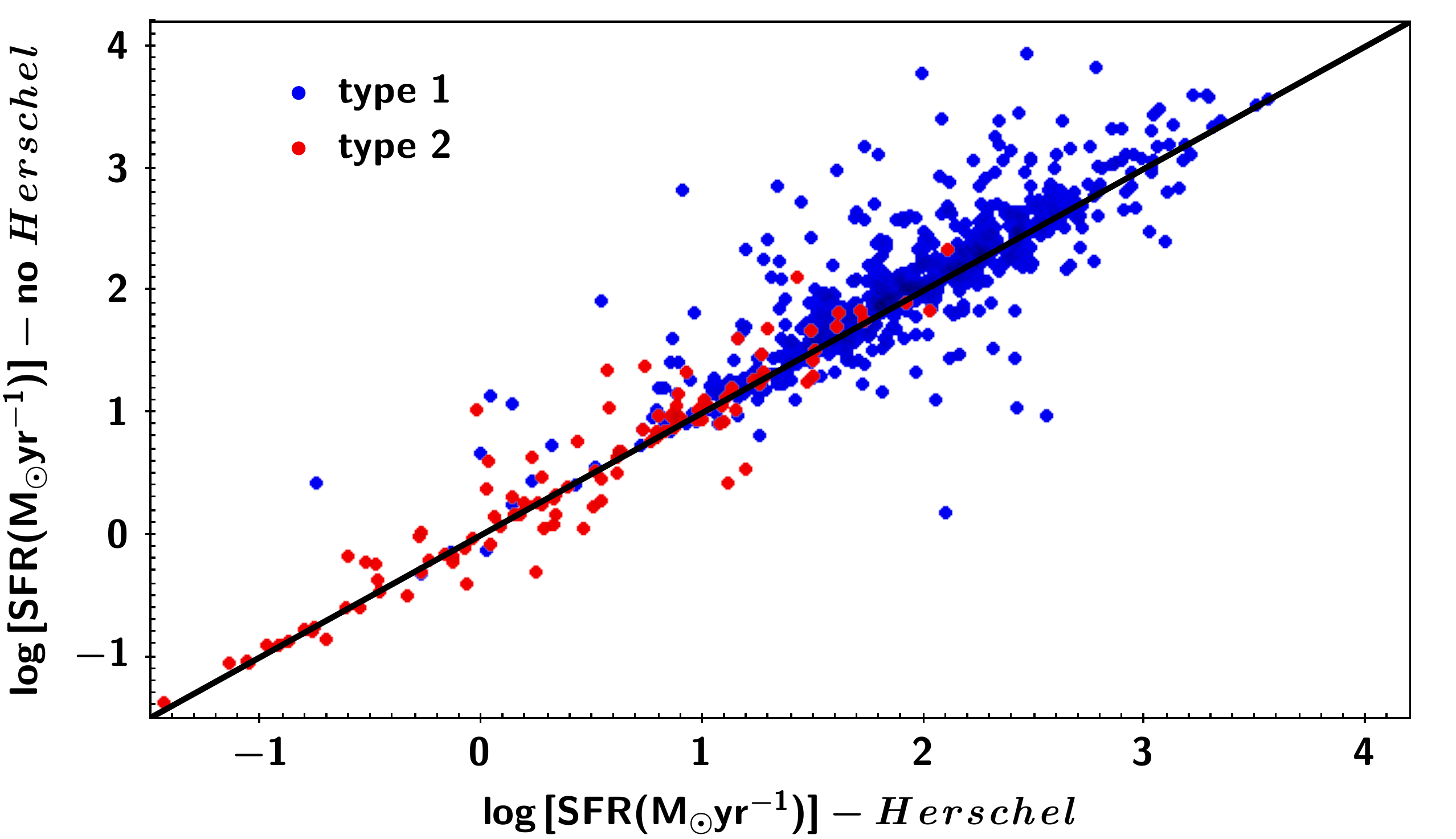}
  \label{}
\caption{Comparison of SFR measurements with and without {\it{Herschel}} photometry, for 683 X-ray sources that have available far-IR photometry. Spectroscopic type 1 sources are shown by blue circles, while type 2 by red circles. Black solid line presents the 1:1 relation. Results show a very good agreement between the two measurements ($\rm SFR_{no\,{\it{Herschel}}}=(0.99\pm0.03)\,SFR_{{\it{Herschel}}}+0.12\pm0.02$). Examining the two AGN types separately, yields, $\rm SFR_{no\,{\it{Herschel}}}=(0.89\pm0.05)\,SFR_{{\it{Herschel}}}+0.23\pm0.05$ and $\rm SFR_{no\,{\it{Herschel}}}=(1.02\pm0.01)\,SFR_{{\it{Herschel}}}+0.02\pm0.01$, for type 1 and 2, respectively.}
\label{fig_herschel_test}
\end{figure}

\begin{figure}
\centering
  \centering
  \includegraphics[width=1.\linewidth, height=7.2cm]{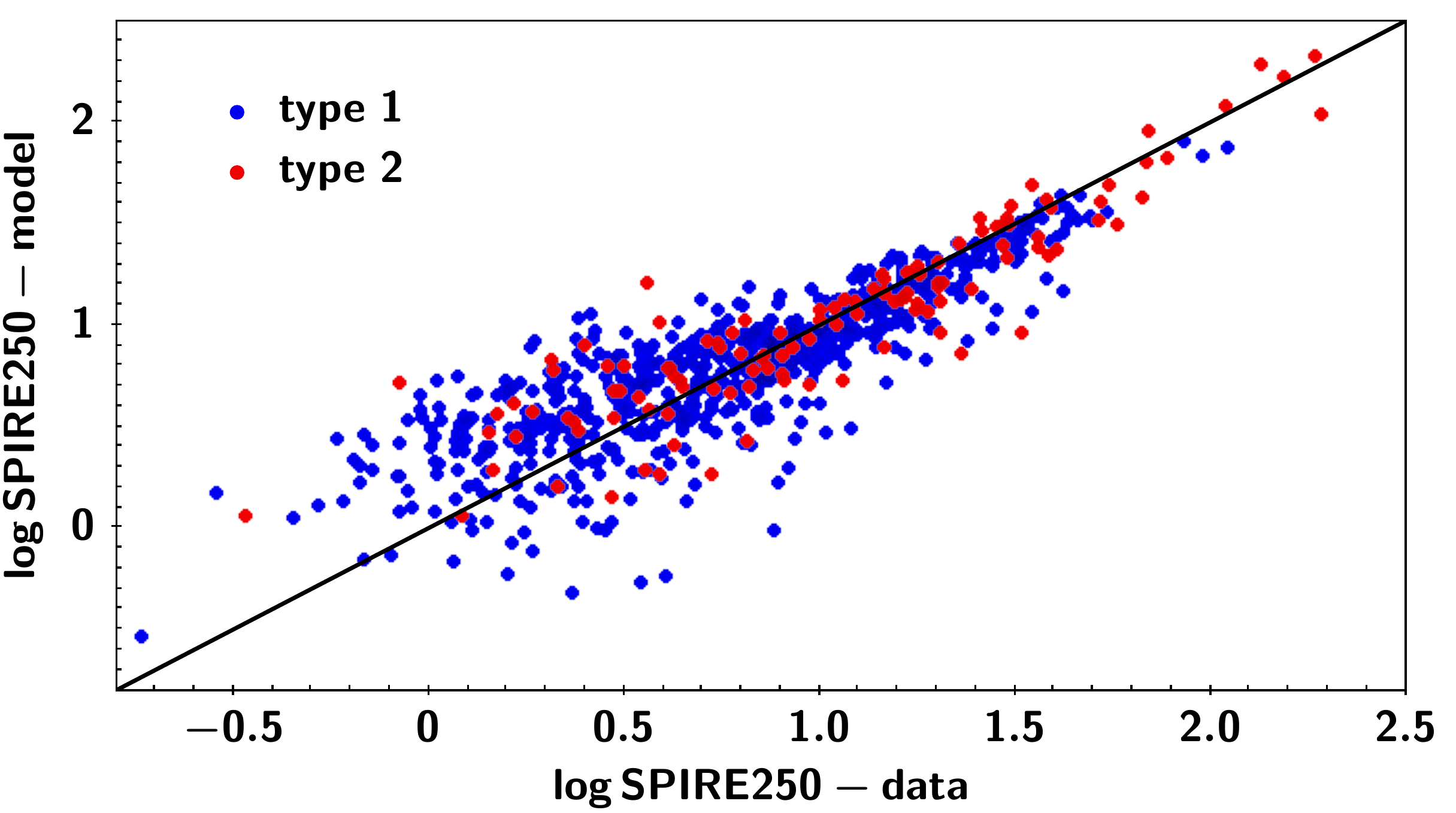}
  \label{}
\caption{Comparison of the values estimated by X-CIGALE with those from the data, for the SPIRE250 flux from {\it{Herschel}}. The mean difference $\rm log\,{\it{Herschel}}_{data} - log\,{\it{Herschel}}_{model}$ is $-0.02$ with a dispersion 0.23, regardless of the AGN type.}
\label{fig_herschel_bands}
\end{figure}

\subsection{The effect of {\it{Herschel}} photometry}
\label{sec_reliability}

683 from the 1201 X-ray sources ($\sim 57\%$) in our sample have been observed by {\it{Herschel}}. For these sources, we run X-CIGALE again, using the same parametric grid described in the previous Section. However, in this second run, we do not take into account the far-IR photometric bands. Our goal is to examine the effect of \it{Herschel}} photometry in the SFR measurements, by comparing the SFR calculations of the two runs. The results are presented in Fig. \ref{fig_herschel_test}. The two measurements are in very good agreement ($\rm SFR_{no\,{\it{Herschel}}}=(0.99\pm0.03)\,SFR_{{\it{Herschel}}}+0.12\pm0.02$). The distribution of the difference $\rm log\,SFR_{no\,{\it{Herschel}}}-log\,SFR_{{\it{Herschel}}}$ has a mean value $\mu =0.10$ and dispersion of 0.33. Examining the results separately for spectroscopic type 1 (blue circles) and type 2 (red circles), yields $\rm SFR_{no\,{\it{Herschel}}}=(0.89\pm0.05)\,SFR_{{\it{Herschel}}}+0.23\pm0.05$ and $\rm SFR_{no\,{\it{Herschel}}}=(1.02\pm0.01)\,SFR_{{\it{Herschel}}}+0.02\pm0.01$, respectively. The distribution of the difference has a mean value of $\mu=0.11, $ and 0.03 with dispersion of 0.35 and 0.26, for type 1 and 2, respectively.

Furthermore, we compare the values estimated by X-CIGALE for the {\it{Herschel}} fluxes with those from the data. The mean difference of $\rm log\,{\it{Herschel}}_{data} - log\,{\it{Herschel}}_{model}$ is $-0.02$, $-0.06$ and 0.08, with a dispersion of 0.23, 0.25, 0.33 for SPIRE250, SPIRE350 and SPIRE500, respectively. The results are similar for both AGN types. In Fig. \ref{fig_herschel_bands}, we present the comparison for SPIRE250. We also compare the reliability of the SFR measurements of these sources that have been observed by {\it{Herschel}} with these that do not have {\it{Herschel}} detection. For that purpose, we use X-CIGALE's ability to create and analyse mock catalogues. These catalogues are based on the best fit model of each source in the dataset. The algorithm uses the best model flux of each source and inserts a noise, extracted from a Gaussian distribution with same standard deviation as the observed flux. Then the mock data are analysed in the same way as the observed data \citep{Boquien2019}. We use the mock catalogues to compare the input and output values of this process and examine the precision of an estimated parameter. The mean difference of the estimated SFR of the mock sources from the input SFR values is 0.01 and 0.00 for {\it{Herschel}} detected and non detected sources, respectively and the corresponding dispersions are 0.23 and 0.26.

Based on these results and the fact that X-ray sources detected by {\it{Herschel}} have similar properties (e.g. redshift, L$_X$) with those non detected by {\it{Herschel}}, we conclude that SFR measurements of those sources in our sample that do not have {\it{Herschel}} photometry, are reliable. \cite{Masoura2018}, found that SFR  calculations without {\it{Herschel}} photometry are systematically underestimated compared to SFR measurements including {\it{Herschel}}. Their X-ray sample consists of sources observed in XXL. However, their dataset includes both spectroscopic and photometric sources, their selection (photometric) criteria are different and different modules and parametric grid was used for the SED fitting.

\begin{figure*}
\centering
\begin{subfigure}{.33\textwidth}
  \centering
  \includegraphics[width=1\linewidth, height=5cm]{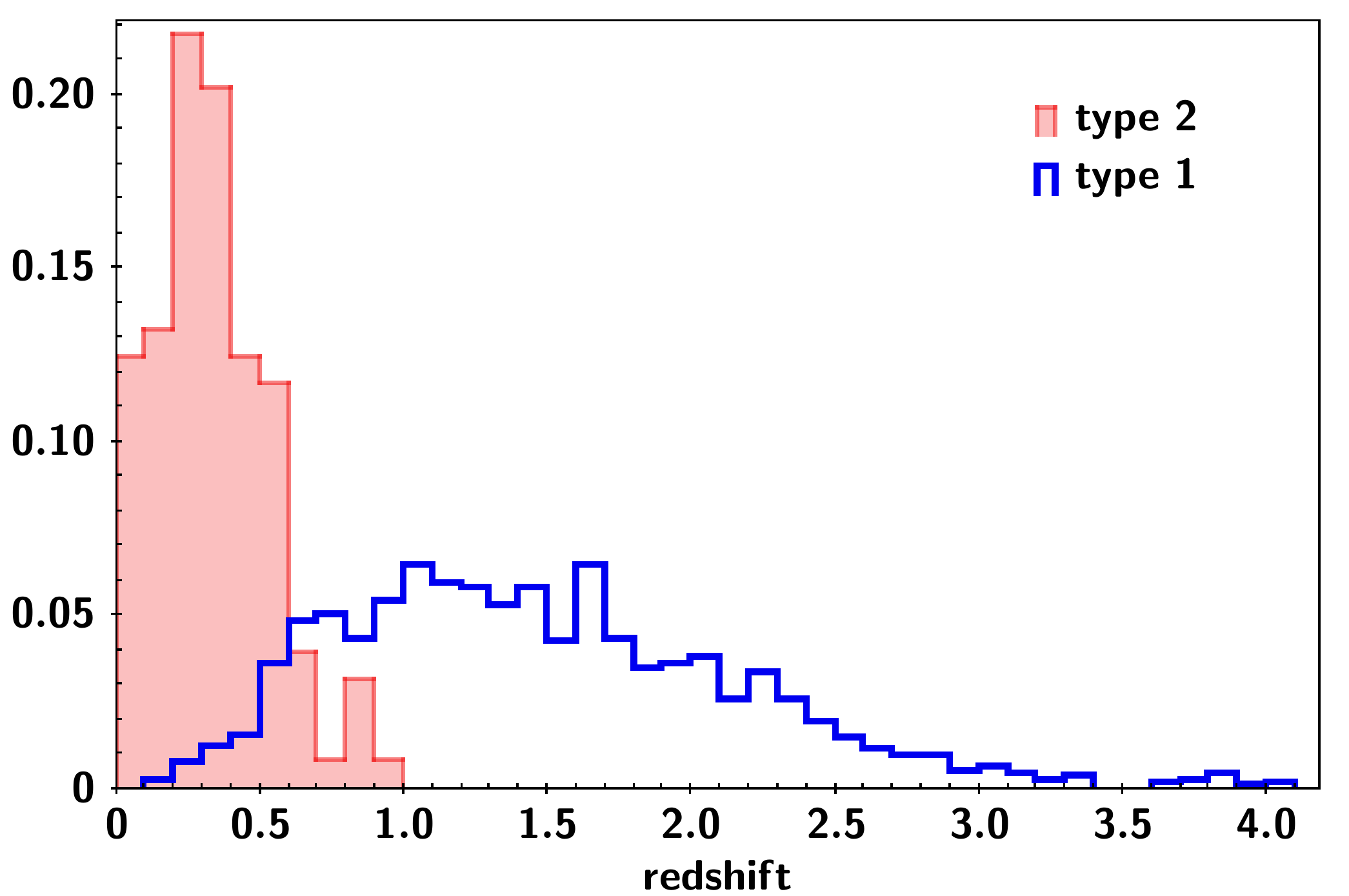}
  \label{}
\end{subfigure}%
\begin{subfigure}{.33\textwidth}
  \centering
  \includegraphics[width=1\linewidth, height=5cm]{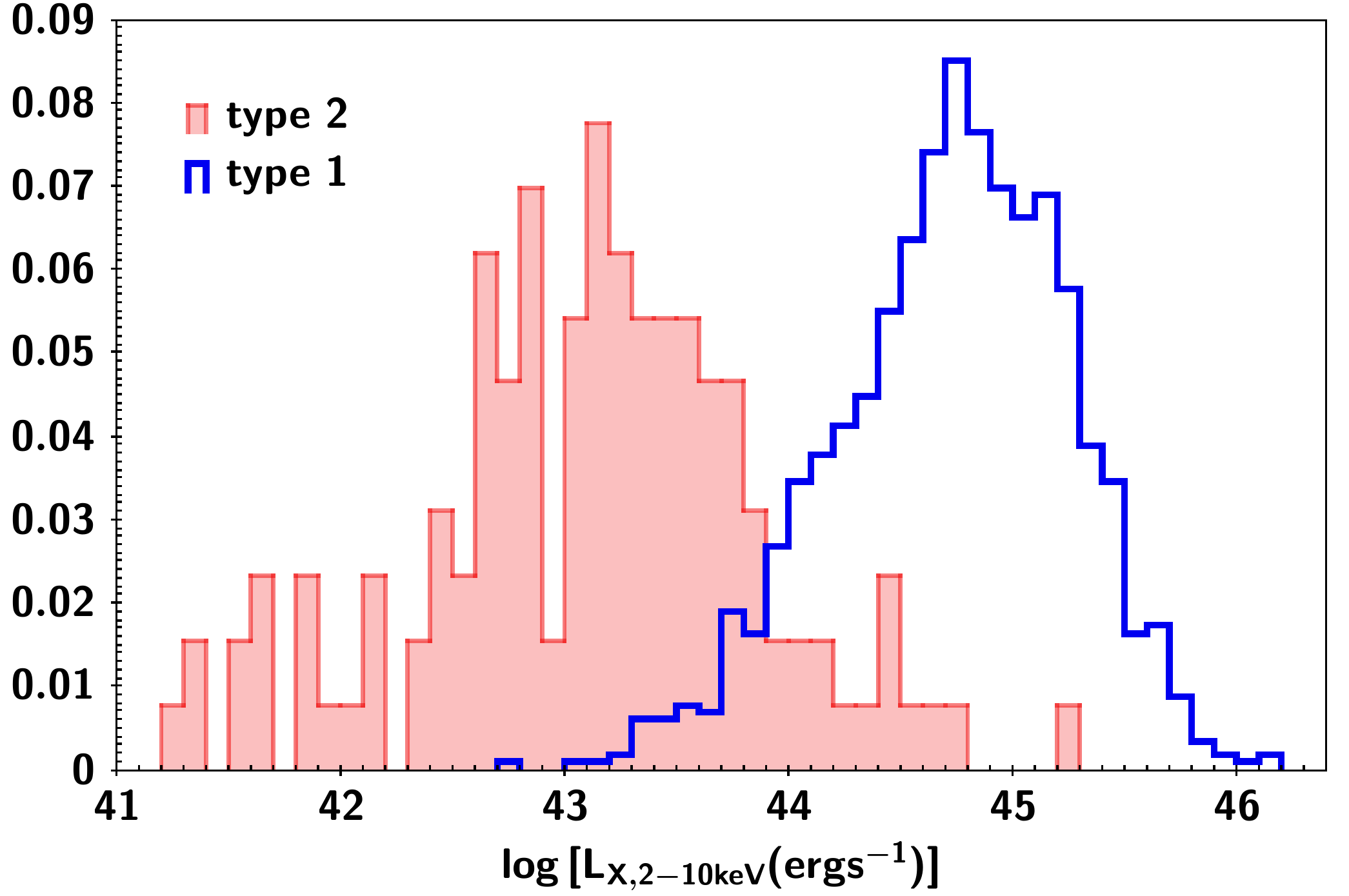}
  \label{}
\end{subfigure}
\begin{subfigure}{.33\textwidth}
  \centering
  \includegraphics[width=1\linewidth, height=5cm]{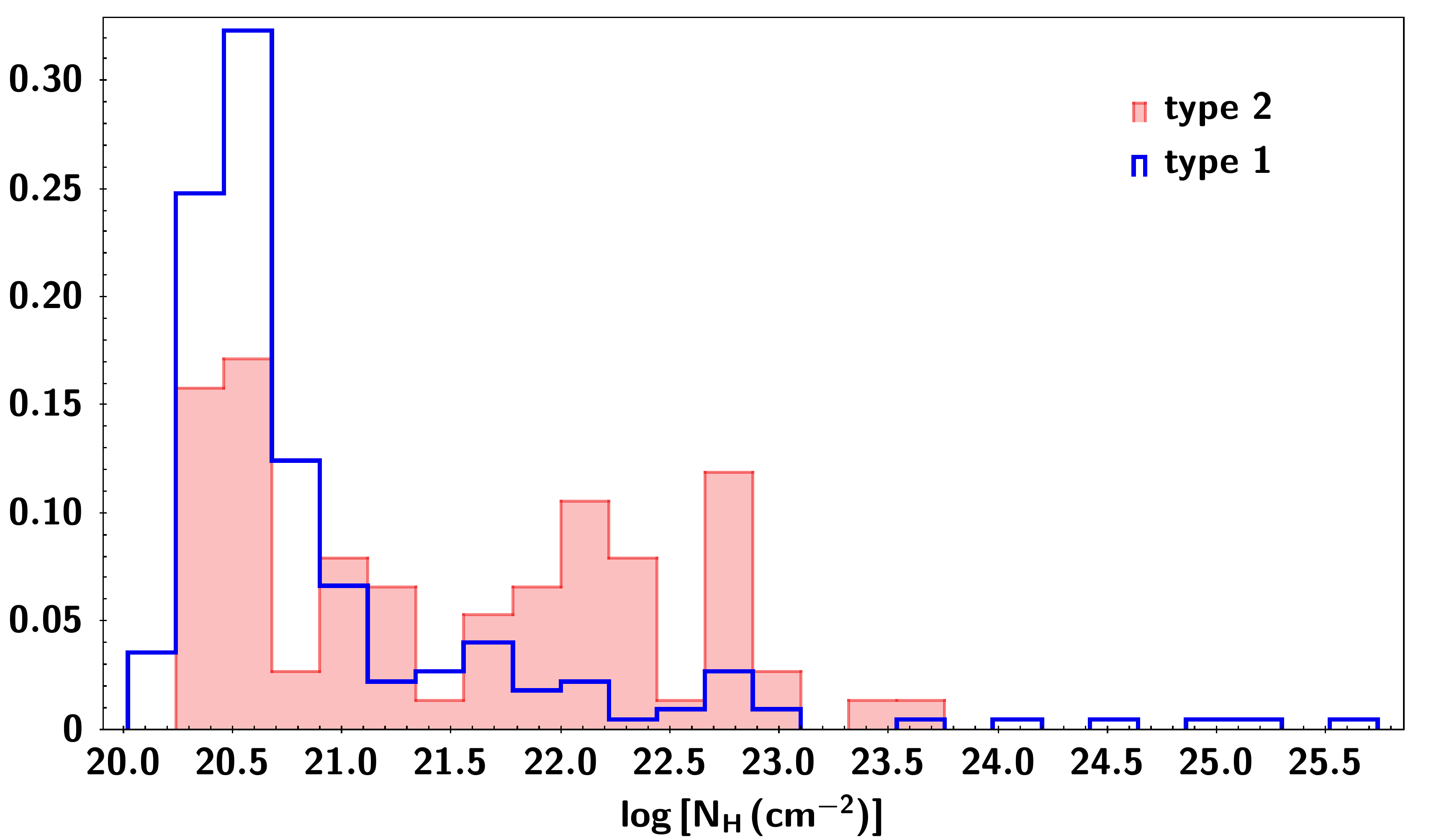}
  \label{}
\end{subfigure}
\caption{Left and middle panels, present the redshift and L$_X$ distributions of type 1 and 2 AGN, among 1201 X-ray sources in our sample. The classification is based on optical spectra. The right panel presents the N$_H$ distribution for the 284 type 1 and 2 X-ray AGN that lie within the same redshift and L$_X$ range.}
\label{fig_type_Lx_redz}
\end{figure*}

\begin{figure*}
\centering
\begin{subfigure}{.505\textwidth}
  \centering
  \includegraphics[width=1\linewidth, height=7.2cm]{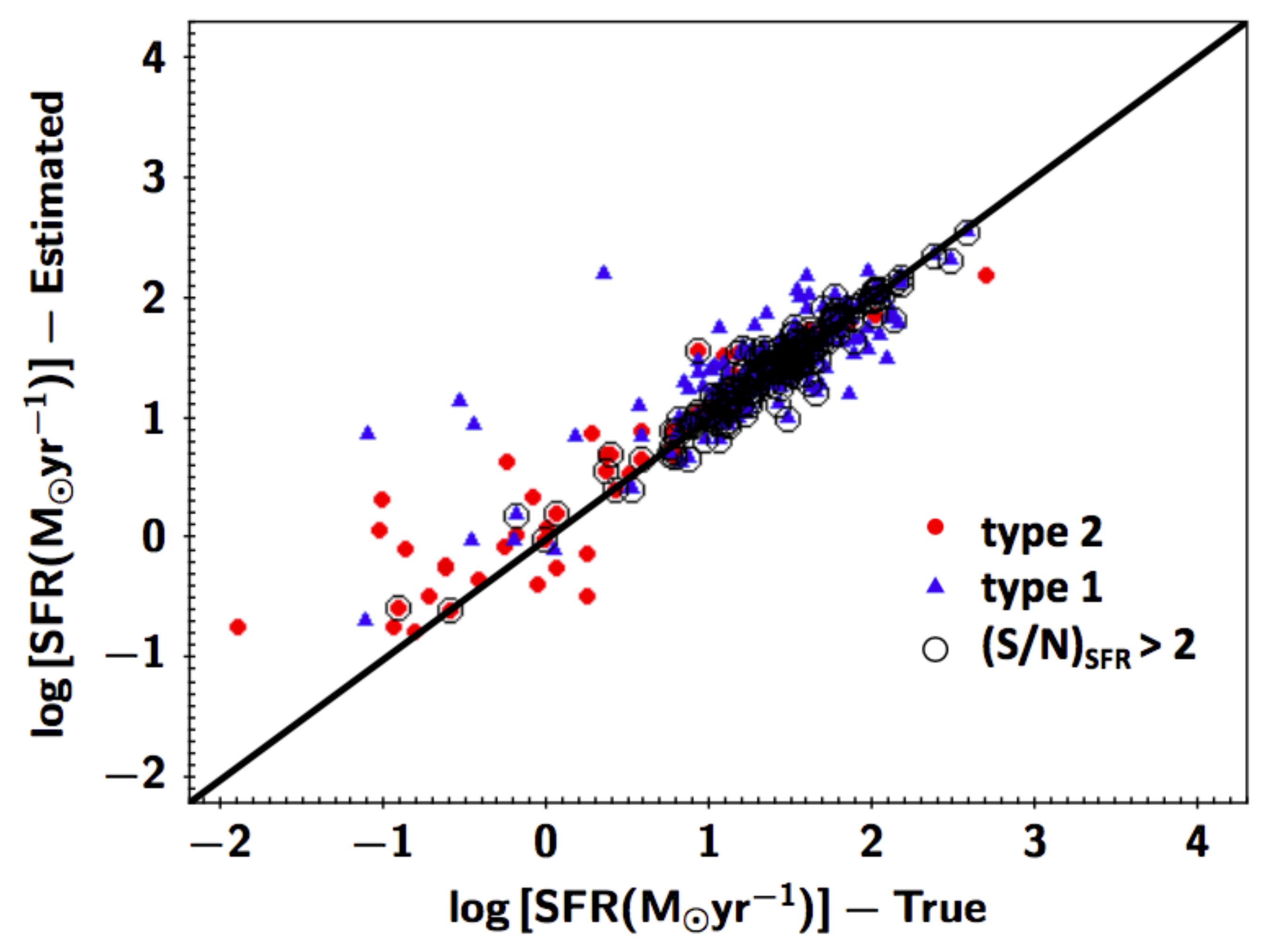}
  \label{}
\end{subfigure}%
\begin{subfigure}{.52\textwidth}
  \centering
  \includegraphics[width=1\linewidth, height=7.2cm]{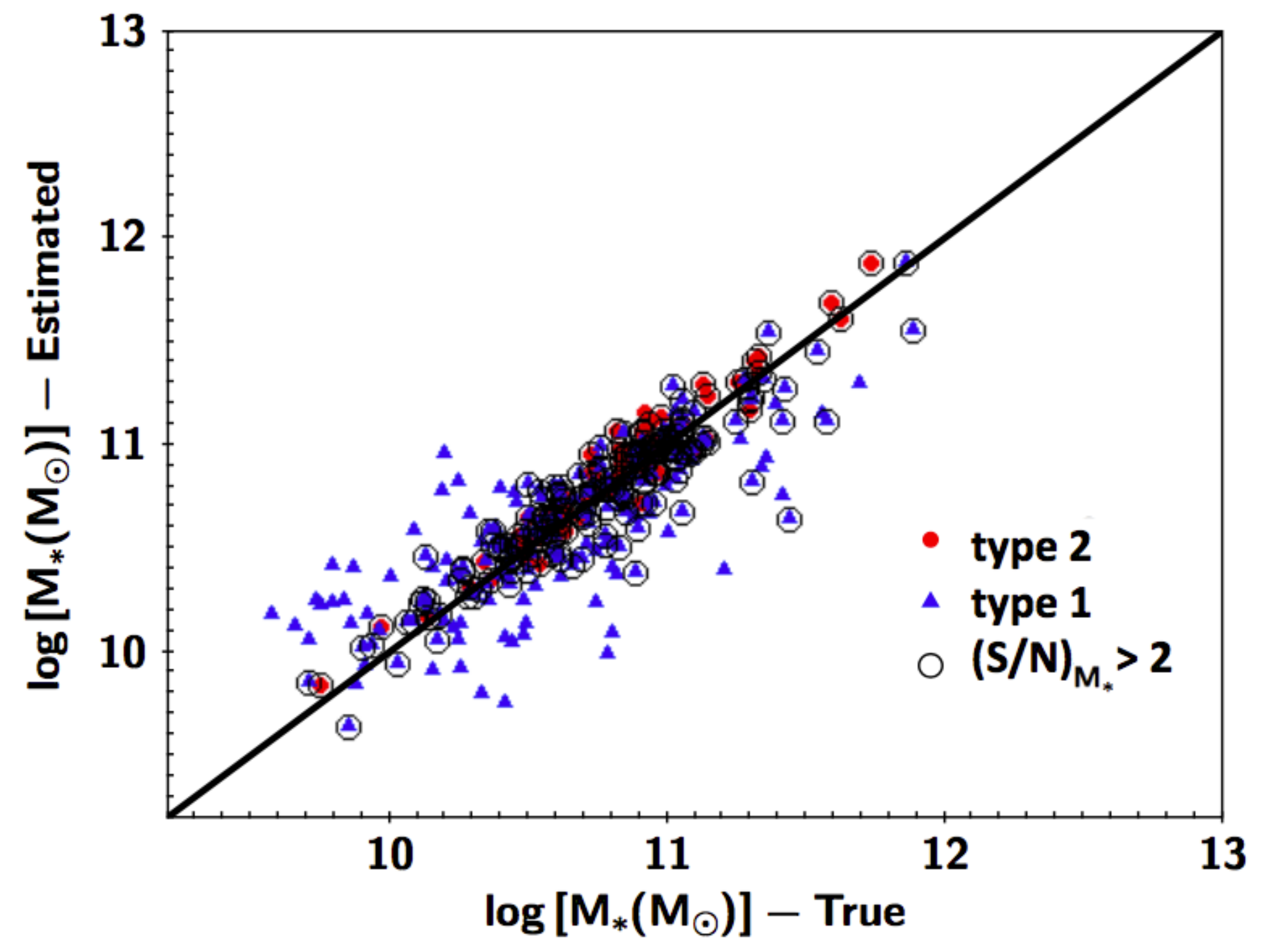}
  \label{}
\end{subfigure}
\caption{Comparison of the SFR and M$_*$ measurements (left and right panel, respectively) for the estimated and true values from the mock analysis. Blue triangles show the results for type 1 X-ray AGN and red circles for type 2. Sources are classified based on optical spectra. The black solid line shows the 1:1 relation. Restricting the measurements to those sources with statistical significance, $\S/N>2$ (open circles), effectively reduces the scatter of the calculations.}
\label{fig_data_mock}
\end{figure*}

\begin{figure*}
\centering
\begin{subfigure}{.33\textwidth}
  \centering
  \includegraphics[width=1\linewidth, height=5.2cm]{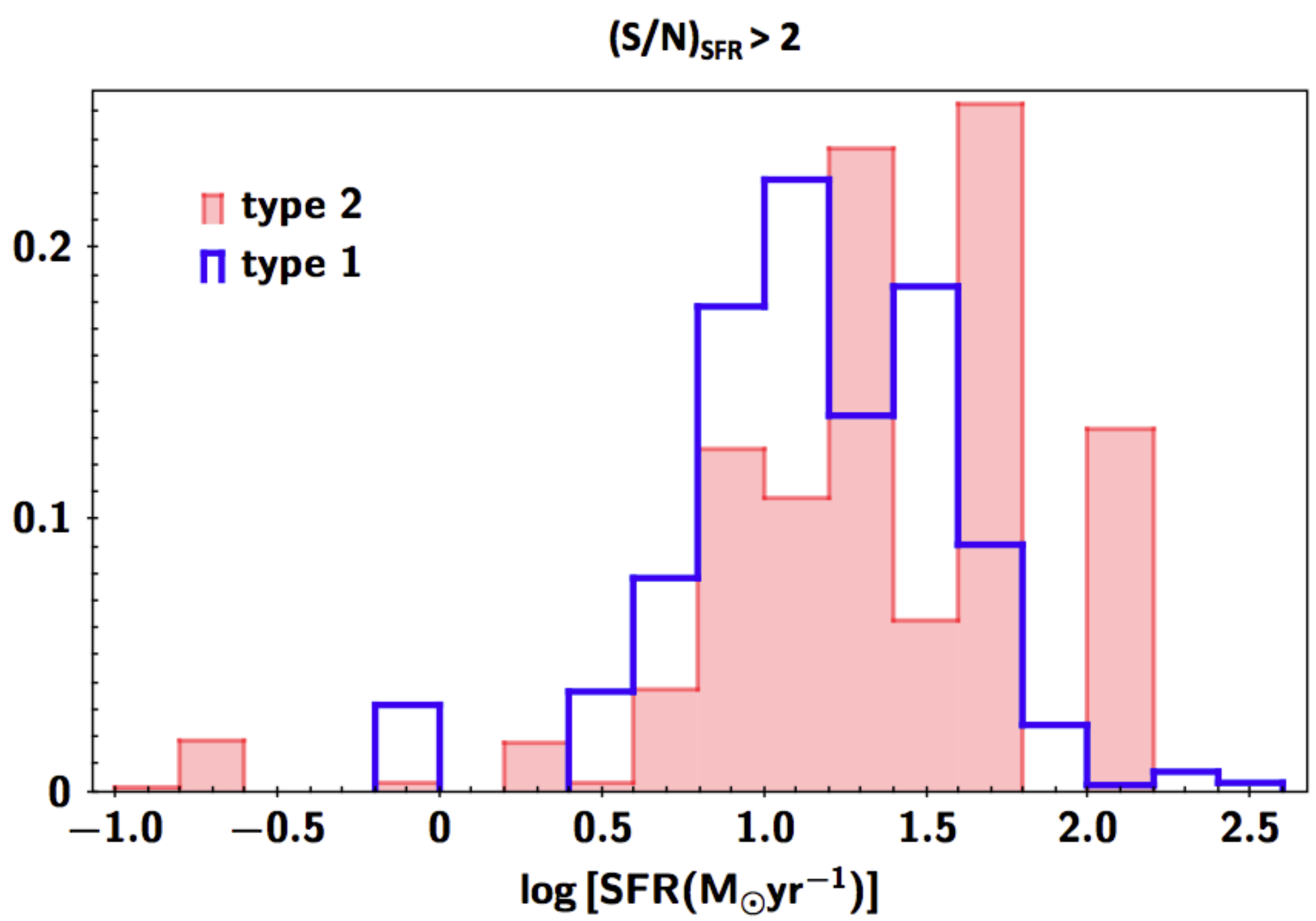}
  \label{}
\end{subfigure}%
\begin{subfigure}{.33\textwidth}
  \centering
  \includegraphics[width=1\linewidth, height=5cm]{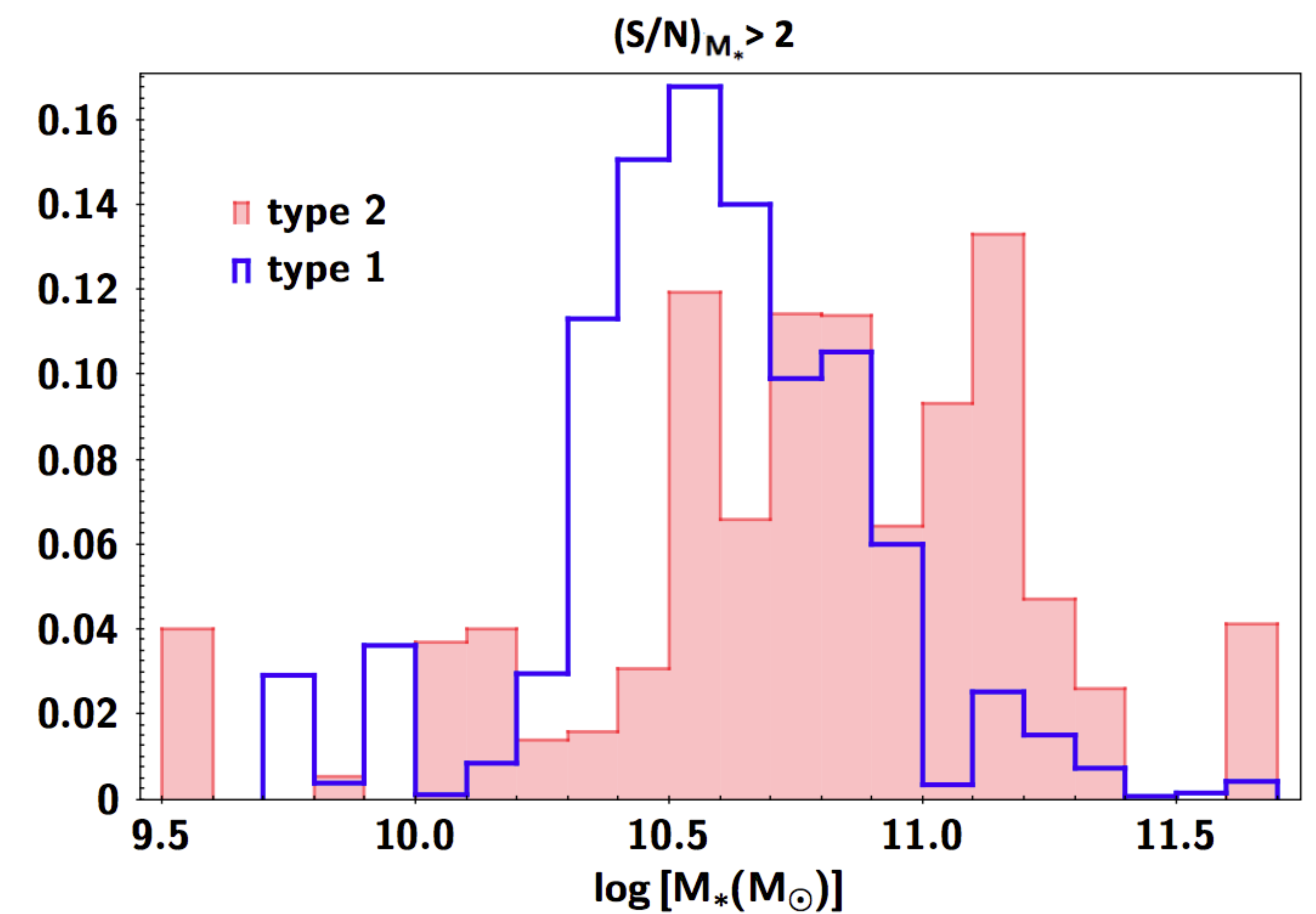}
  \label{}
\end{subfigure}
\begin{subfigure}{.33\textwidth}
  \centering
  \includegraphics[width=1.\linewidth, height=5cm]{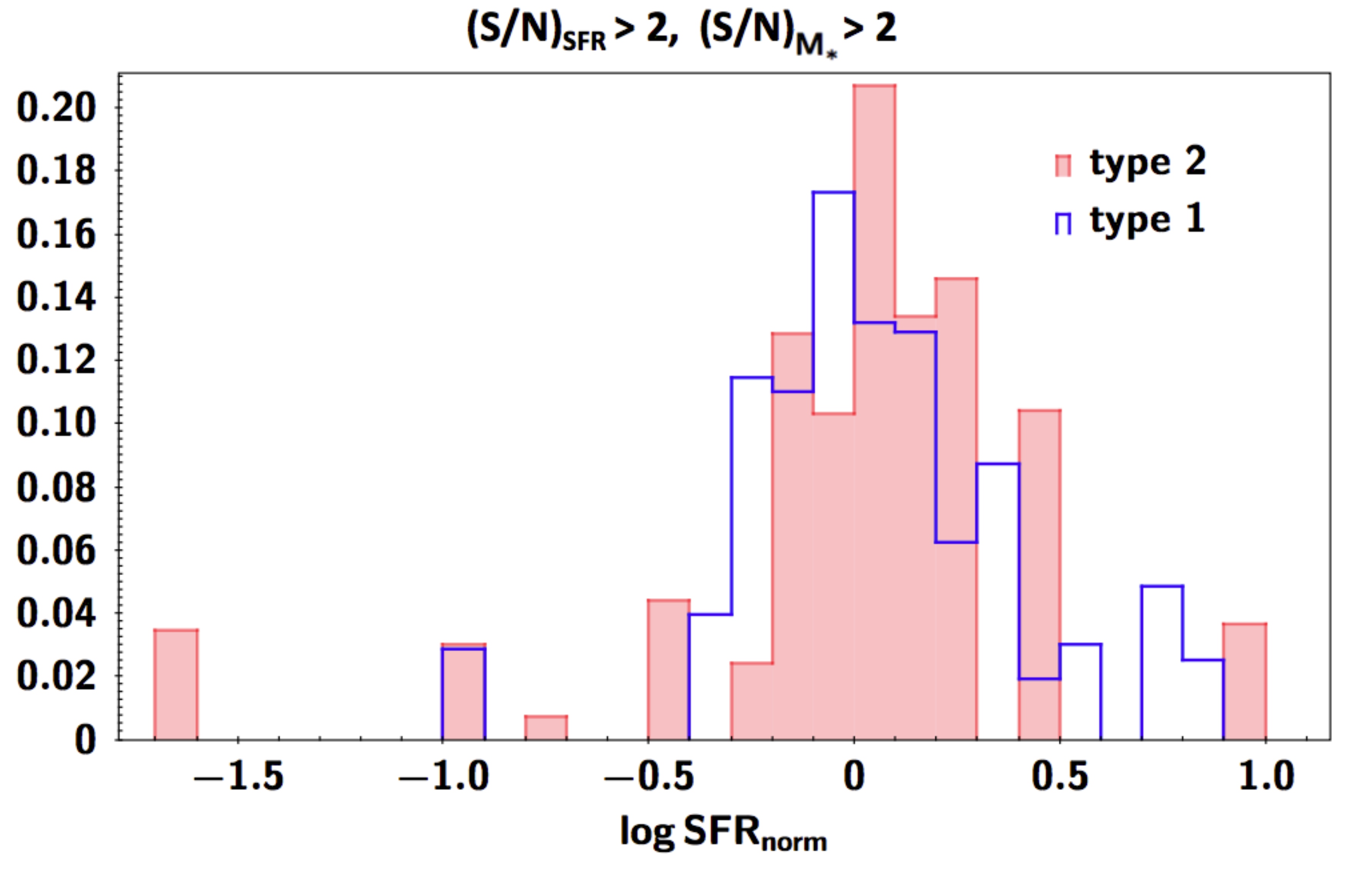}
  \label{}
\end{subfigure}
\caption{From left to right: Star formation rate, stellar mass and SFR$_{norm}$ distributions of type 1 (blue lines) and type 2 (red shaded histograms) X-ray selected AGN. The classification is based on optical spectra. SFR and SFR$_{norm}$ distributions of the two populations are similar. However, type 2 AGN tend to reside in galaxies with stellar mass $\sim 0.30$\,dex higher than type 1 AGN ($\rm log\,[M_*(M_\odot)]=10.87^{+0.06}_{-0.12}$ and $10.57^{+0.20}_{-0.12}$, respectively). Although, this difference is statistical significant only at $\approx 1\sigma$, is in agreement with previous studies.}
\label{fig_type_host}
\end{figure*}

\section{Host galaxy properties of obscured and unobscured AGN classified based on optical spectra}
\label{sec_optspectra}

In this Section, we estimate the host galaxy properties of spectroscopic type 1 and 2 X-ray AGN. We select those sources with the most robust measurements and then compare the SFR, M$_*$ and SFR$_{norm}$ of the two AGN populations. SFR$_{norm}$ is defined as the ratio of the SFR of AGN to the SFR of star-forming main sequence (MS) galaxies with the same stellar mass and redshift \citep[e.g.,][]{Mullaney2015, Masoura2018, Bernhard2019, Masoura2021}. The \cite{Schreiber2015} analytical formula is used for the calculation of the latter.

\subsection{Properties of type 1 and type 2 X-ray AGN}

For the estimation of SFR and M$_*$, we apply SED fitting using the grid described in Section \ref{sec_cigale}. In this case, the inclination angle is fixed to the value that corresponds to the AGN type indicated by the optical spectra, i.e., $30^\circ$ for type 1 and $70^\circ$ for type 2. In addition to the SFR and M$_*$ properties, we also compare the SFR$_{norm}$ parameter of the two AGN types. For the latter, we use expression (9) of \cite{Schreiber2015}. As mentioned in Mountrichas et al. 2021b, using an expression from the literature to calculate the SFR of star-forming MS galaxies and compare it with the SFR of X-ray AGN to estimate SFR$_{norm}$, hints at a number of systematics. For example, different methods and/or (SED fitting) algorithms may be used in the estimation of SFR and different definitions of MS may be applied in different studies. Thus, we only wish to examine in a qualitative manner whether the SFR$_{norm}$ parameter differs for different AGN classifications and not draw conclusions regarding the position of the two AGN types relative to the MS.

We split the X-ray sample into type 1 and type 2 sources, using optical spectra. This information is available in the public catalogue presented in \cite{Menzel2016}. The classification rules are described in detail in Sections 2 and 3 of Menzel et al. In brief, the optical spectroscopic follow-up was performed using the {\it{BOSS}} spectrograph of SDSS \citep{Smee2013}. The full width at half maximum (FWHM) was estimated for emission lines originating from different regions of the AGN (H$\beta$, MgII, CIII and CIV). A source is classified as type 1 when an emission line has FWHM larger than 1000\,$\rm km\,s^{-1}$. In our analysis, we only use AGN with reliable spectral classification, e.g. sources with lines that have low significance due to very strong host galaxy continuum contribution or very low signal to noise ratio spectra, are classified as type 2 candidates in the Menzel et al. catalogue (NLAGN2cand) and are excluded from our analysis. Therefore, we consider only sources that are labeled as BLAGN1 (type 1) and NLAGN2 (type 2), in the catalogue of \cite{Menzel2016}. From the 1,201 X-ray sources in our dataset (see Section \ref{sec_unreliable}), 1,109 are type 1 and 92 are type 2. 

Redshift and X-ray luminosity distributions of the two populations are presented in Fig. \ref{fig_type_Lx_redz}. We notice that there are no type 2 sources above $\rm z>1$. This is due to two effects. It is well known that the fraction of type 2 AGN decreases at higher luminosities \citep[e.g.][]{Merloni2014, Aird2015}. Moreover, at $\rm z>1.0$, the magnitude of type 2 sources reaches $\rm r_{model}=22.5$\,mag and the host galaxy becomes too faint to be detected by SDSS images \citep{Menzel2016}. On the other hand, spectroscopic type 1 AGN are biased towards high luminosity sources, since broad lines are harder to detect at lower luminosities. To compare the host galaxy properties of the two subsamples, we first account for their different L$_X$ and redshift distributions. For that purpose, a weight is assigned to each source. For the calculation of the weight, we join the redshift distributions and normalise each one by the total number of sources in each redshift bin (in bins of 0.1). The same process is repeated for the L$_X$ distributions (in bins of 0.1\,dex). This gives us the PDF in this 2-D (L$_X$, redshift) space. The latter is used to weigh each source based on its redshift and X-ray luminosity \citep[e.g.][]{Mountrichas2016, Masoura2021, Mountrichas2021b}. In practice, this means that we restrict the redshift range of type 1 and 2 sources to $\rm 0.1<z<0.9$ and the X-ray luminosity  to $43.0<\rm log\,[L_{X,2-10keV}(ergs^{-1})]<44.8$. Or equally, that sources outside these redshift and luminosity ranges are assigned a zero weight. There are 284 X-ray AGN that satisfy these conditions. 220 are type 1 and 64 are type 2.

The hydrogen column density, N$_H$, quantifies the X-ray absorption of a source. This parameter has been estimated for all AGN in the Menzel et al. catalogue, in \cite{Liu2016}. N$_H$ have been calculated, by applying X-ray spectral modelling and stacking, adopting the Bayesian X-ray Analysis software \citep[BXA;][]{Buchner2014} to fit the X-ray spectra of individual sources. The right panel of Fig. \ref{fig_type_Lx_redz}, presents the N$_H$ distributions of the two AGN types for the 284 AGN. The vast majority ($90\%$) of type 1 sources are X-ray unabsorbed (${\rm{N}}_H<10^{22}\,\rm cm^{-2}$). On the other hand, optically classified type 2 AGN have a broad range of N$_H$, split nearly equal between X-ray absorbed and unabsorbed sources. These trends \citep[also seen in Fig. 9 of][]{Liu2016} are consistent with those found in XMM-COSMOS \citep{Merloni2014} and XMM-LSS \citep[][see also \cite{Trouille2009}]{Garcet2007}. N$_H$ measurements become less accurate for X-ray spectra with low number of photons. Restricting the comparison of the N$_H$ distributions to sources with $>50$ net photons does not change the results. This is also true if we replace the N$_H$ values from the Liu et al. catalogue with those estimated in \cite{Mountrichas2021} that calculated N$_H$, via hardness ratios, applying a Bayesian approach \citep{Park2006}. A possible scenario is that the observed trends are related to the large scatter that X-ray and optical obscuration present. This is due to e.g. X-ray column density variability \citep{Yang2016}, obscuration located at galactic scales that is not related to nuclear absorption \citep{Malizia2020} and a complex AGN structure \citep{Ogawa2021, Arredondo2021}. There are studies that found X-ray absorbed sources with broad UV/optical lines \citep[e.g.][]{Li2019} and optical type 2 sources that have low X-ray absorption \citep[e.g.][]{Masoura2020}. 

\subsection{Comparison of host galaxy properties of type 1 and 2 AGN}
\label{sec_comparison_host_results}

Estimation of host galaxy properties are more challenging for galaxies that host AGN compared to non-AGN systems. This is especially true for unobscured AGN. The AGN emission can outshine the optical emission of the host galaxy, thus increasing the uncertainties of the measurements. To estimate SFR, the SED code is, additionally, required to disentangle the IR emission of the host galaxy from the IR emission of the AGN. To examine the impact of these effects on the reliability of the SFR and M$_*$ measurements, we use the mock catalogues created by X-CIGALE.  Fig. \ref{fig_data_mock} presents the results for the SFR (left panel) and M$_*$ (right panel) calculations, for the 284 X-ray AGN. The vertical axis presents the Bayesian values of the parameter obtained from the fit of the mock sources while the horizontal axis presents the true (input) values from the best fit of the data. Although, X-CIGALE recovers successfully the true SFR and M$_*$ values for most of the X-ray sources, there is a scatter, in particular in the M$_*$ measurements of type 1 AGN. To reduce this scatter, we estimate the statistical significance of each measurement and consider in our analysis only sources with the most robust measurements ($S/N>2$). These sources are marked with an open circle in Fig. \ref{fig_data_mock}. Specifically, for the comparison of SFR between different AGN types we keep only sources with $(S/N) _{SFR}>2$. Similarly, only sources with $(S/N) _{M_*}>2$ are included in the comparison of the stellar mass distributions of the two populations. In the case of SFR$_{norm}$, both requirements are applied. This effectively reduces the scatter of the SFR and M$_*$ measurements. Table \ref{table_numbers}, presents the number of sources that satisfy the aforementioned criteria. 



\begin{table*}
\caption{Number of X-ray sources classified as type 1 and type 2, using optical spectra and SED fitting, that have reliable SFR and M$_*$ measurements.}
\centering
\setlength{\tabcolsep}{3.mm}
\begin{tabular}{ccccccc}
classification criterion & \multicolumn{2}{c}{$(S/N) _{SFR}>2$} & \multicolumn{2}{c}{$(S/N) _{M_*}>2$}  & \multicolumn{2}{c}{$(S/N)_{SFR}>2$ \& $(S/N)_{M_*}>2$}\\
\hline
 & type 1 & type 2 & type 1 & type 2 & type 1 & type 2 \\
\hline
optical spectra & 142 & 42 & 132 & 70 & 89 & 39 \\
X-CIGALE & 430 & 257 & 284 & 292 & 229 & 246 \\
\label{table_numbers}
\end{tabular}
\tablefoot{The reliability is quantified by the statistical significance, $S/N$ (see text for more details). The numbers for the type 1 and 2 AGN refer, to the subsample of 284 sources that lie within the same redshift and L$_X$ range. The numbers for the X-CIGALE classification refer to these sources, among  the 972 AGN (out of the 1291) with secure classification from the SED fitting analysis that satisfy the $(S/N)>2$ criterion (see text for more details). }
\end{table*}

\begin{table*}
\caption{Mean values and errors, estimated via bootstrap resampling, for the SFR, M$_*$ and SFR$_{norm}$ distributions of spectroscopic type 1 and 2 X-ray AGN, presented in Fig. \ref{fig_type_host}.}
\centering
\setlength{\tabcolsep}{3.mm}
\begin{tabular}{cccccc}
\hline
\multicolumn{2}{c}{$\rm log\,SFR$} & \multicolumn{2}{c}{$\rm log\,M_*$}  & \multicolumn{2}{c}{$\rm log\, SFR_{{\it{norm}}}$}\\
\multicolumn{2}{c}{$\rm (M_\odot yr^{-1})$} & \multicolumn{2}{c}{($\rm M_\odot)$}  & \multicolumn{2}{c}{}\\
\hline
type 1 & type 2 & type 1 & type 2 & type 1 & type 2 \\
\hline
$1.27^{+0.15}_{-0.11}$ & $1.36^{+0.18}_{-0.22}$ & $10.57^{+0.20}_{-0.12}$ & $10.87^{+0.06}_{-0.12}$ & $0.03^{+0.08}_{-0.07}$ & $0.14^{+0.05}_{-0.12}$ \\
\label{table_resampling}
\end{tabular}
\end{table*}

\begin{figure*}
\centering
\begin{subfigure}{.33\textwidth}
  \centering
  \includegraphics[width=1\linewidth, height=5cm]{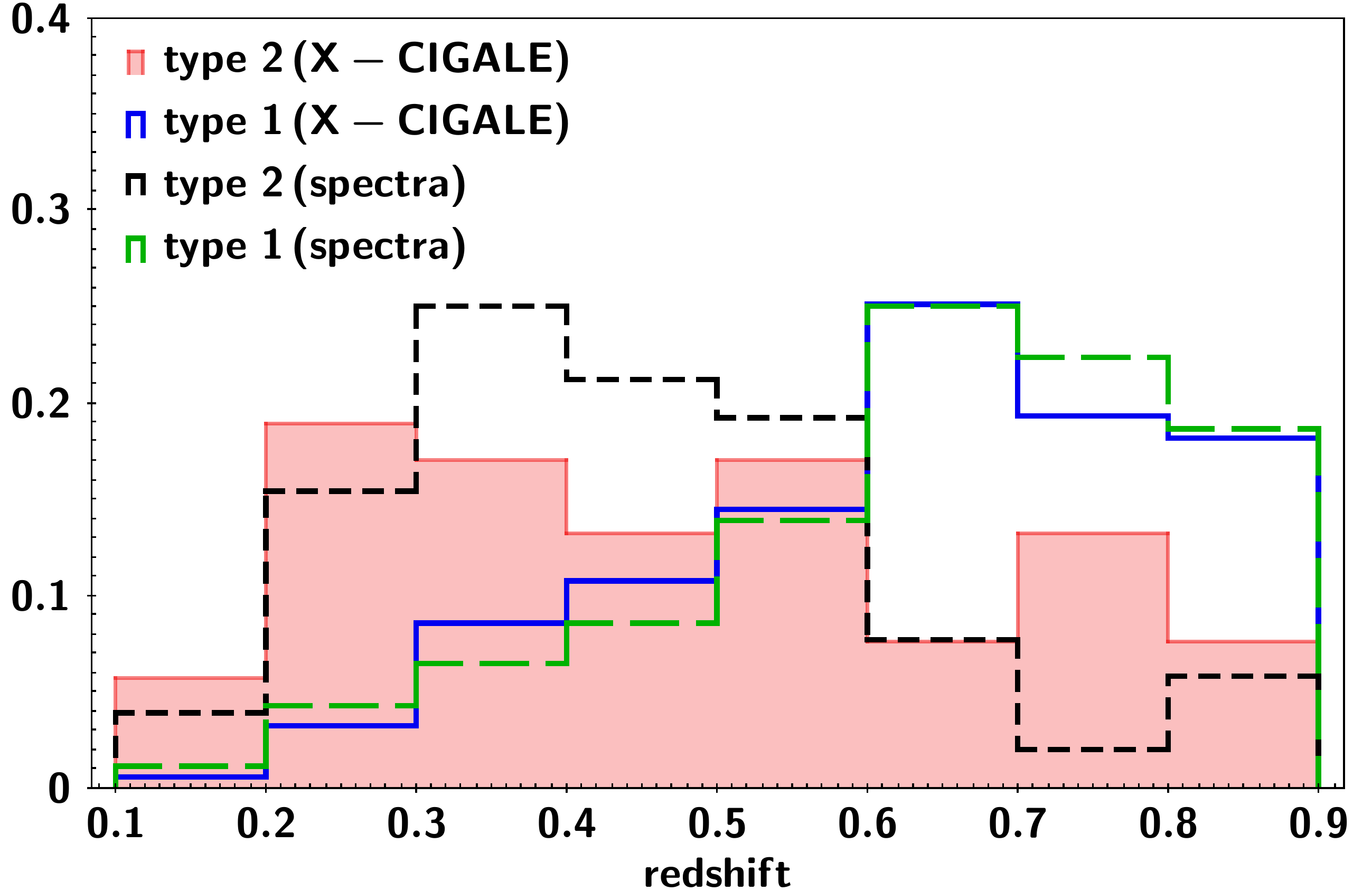}
  \label{}
\end{subfigure}%
\begin{subfigure}{.33\textwidth}
  \centering
  \includegraphics[width=1\linewidth, height=5cm]{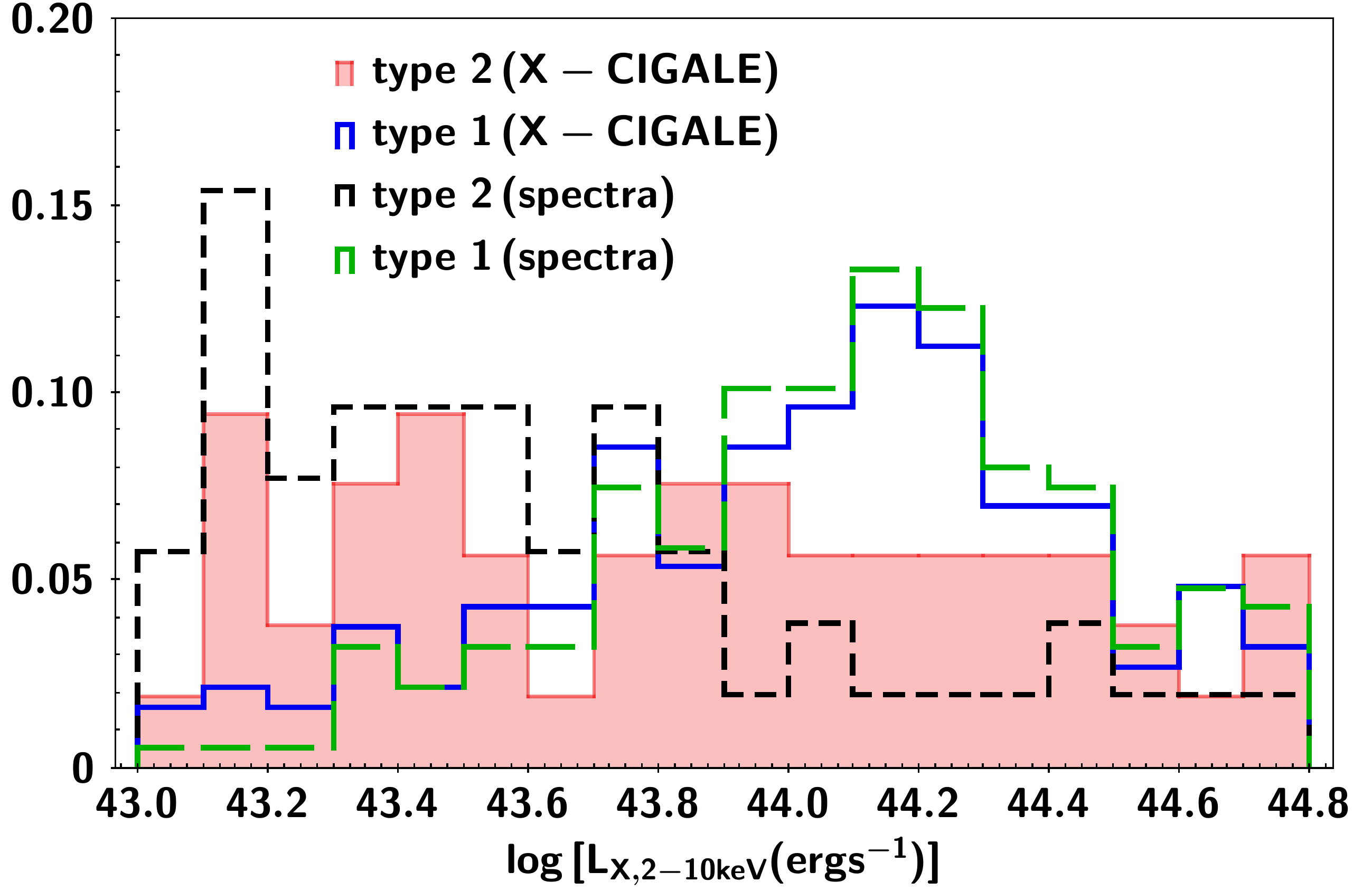}
  \label{}
\end{subfigure}
\begin{subfigure}{.33\textwidth}
  \centering
  \includegraphics[width=1\linewidth, height=5cm]{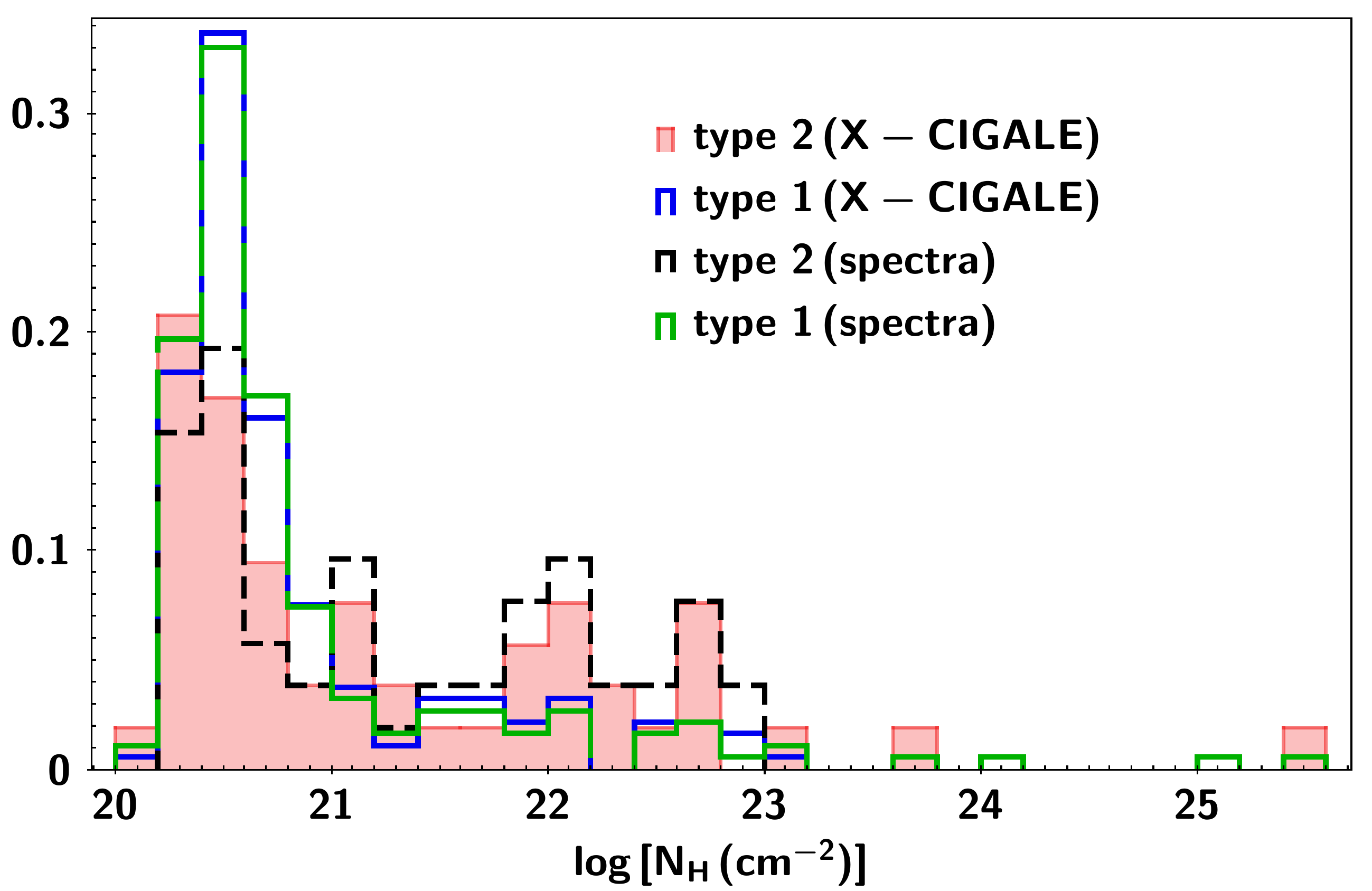}
  \label{}
\end{subfigure}
\caption{Redshift, L$_X$ and N$_H$ distributions of type 2 and type 1 AGN, based on the inclination angle estimated by X-CIGALE (red shaded area and blue line respectively). For comparison, the distributions of type 1 and 2 sources, based on optical spectra, are also presented, with green and black dashed lines, respectively.}
\label{fig_cigale_Lx_redz}
\end{figure*}

\begin{figure*}
\centering
  \centering
  \includegraphics[width=1.\linewidth, height=7.2cm]{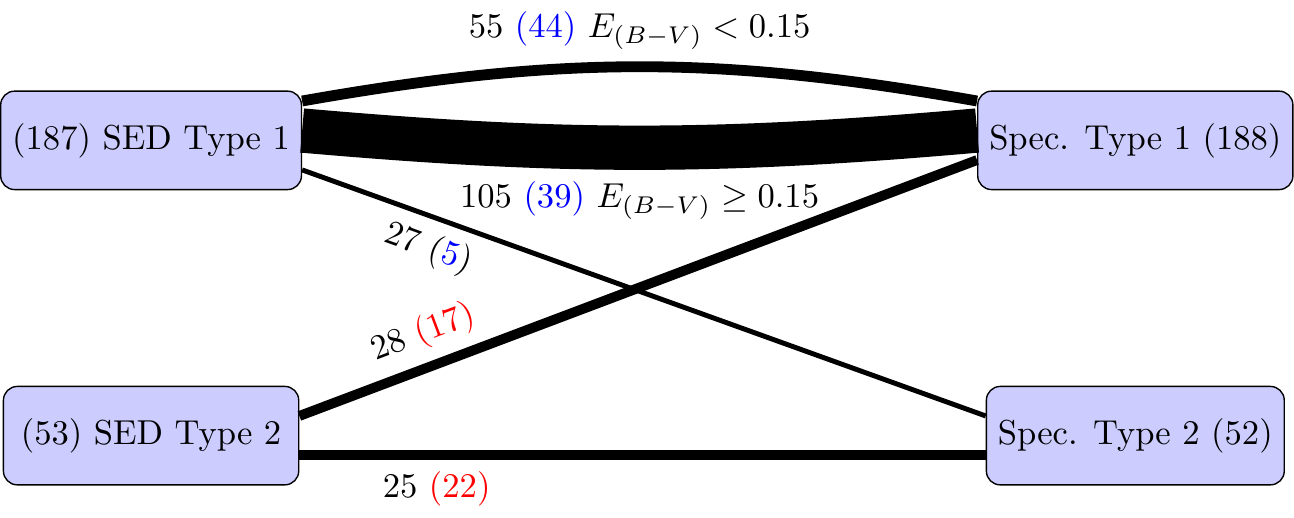}
  \label{}
\caption{Comparison of the classification of the 240 X-ray AGN in our dataset, using different classification criteria. The left side presents the number of sources classified based on SED fitting, while the right side shows the number of sources classified based on optical spectra. Numbers in the parentheses present the classification using the \citep{Hickox2017} criterion (the number of red sources is shown in red and non red sources in blue).}
\label{fig_comp_referee}
\end{figure*}

SFR, stellar mass and SFR$_{norm}$ distributions of type 1 and 2 X-ray AGN are presented in Fig. \ref{fig_type_host}. We note, that for the comparison of the galaxy properties, the redshift and X-ray luminosity distributions of the various type 1 and 2 AGN subsamples have been matched, by weighting each source, following the procedure described above. The two-sample Kolmogorov-Smirnov test (KS-test) and the Mann-Whitney test (MW-test), show that the SFR distributions (left panel) of the two populations are similar. The $p-\rm values$ are 0.77 and 0.47, respectively. The two tests, show that the SFR$_{norm}$ distributions are also similar (right panel; $p-\rm values=0.82, 0.56$, for the KS- and MW-test, respectively). However, the M$_*$ distribution of type 2 X-ray AGN peaks at higher M$_*$ values, compared to their type 1 counterparts (middle panel). The difference is not statistically significant based on the KS- and MW-tests ($p-\rm values=0.24, 0.17$, respectively). We estimate the errors on the mean values, using bootstrap resampling \citep[e.g.,][]{Loh2008}. The results are presented in table \ref{table_resampling}. Based on these measurements, the difference in stellar mass of type 1 and 2 X-ray AGN is $\sim 0.3$\,dex, but is significant only at $1\,\sigma$. However, this difference is in agreement with previous studies. \cite{Zou2019}, used optical spectra, morphology and optical variability to classify X-ray sources in the COSMOS field into type 1 and type 2. Their results showed no difference in the SFR distributions of the two populations. However, they found that type 1 AGN tend to live in galaxies with smaller, by 0.2\,dex, stellar mass than their type 2 counterparts, at a significance of $\approx 4\sigma$. Thus, both our study and \cite{Zou2019} find a very similar difference in the mean stellar mass of galaxies that host type 1 vs. type 2 X-ray AGN. A possible reason that the result of \cite{Zou2019} has a higher statistical significance than ours is the additional criteria they applied to classify sources.


\cite{Zou2019} suggested that the small difference in stellar mass, between type 1 and type 2 AGN is related to the lower X-ray absorption of type 1 sources. Some previous studies found a correlation between N$_H$ and M$_*$ \citep{Buchner2017, Lanzuisi2017}. However, other X-ray studies found that X-ray absorbed and unabsorbed AGN live in galaxies with similar stellar mass \citep{Merloni2014, Masoura2021, Mountrichas2021b}. In our sample, most type 1 sources have lower N$_H$ values (median of $\rm log\,[N_H\,(cm^{-2})=20.60$) compared to type 2 sources (median of $\rm log\,[N_H\,(cm^{-2})]=21.49$), but the distribution of the latter is very broad (Fig. \ref{fig_type_Lx_redz}).

We conclude that AGN classified as type 1 and 2 based on optical spectra, have similar SFR and SFR$_{norm}$ distributions. On the other hand, type 2 AGN live, on average, in more massive galaxies than their type 1 counterparts. Although this difference does not appear statistical significant is in agreement with previous studies.

\begin{figure}
\centering
\begin{subfigure}[b]{0.5\textwidth}
   \includegraphics[width=1\linewidth, height=7.2cm]{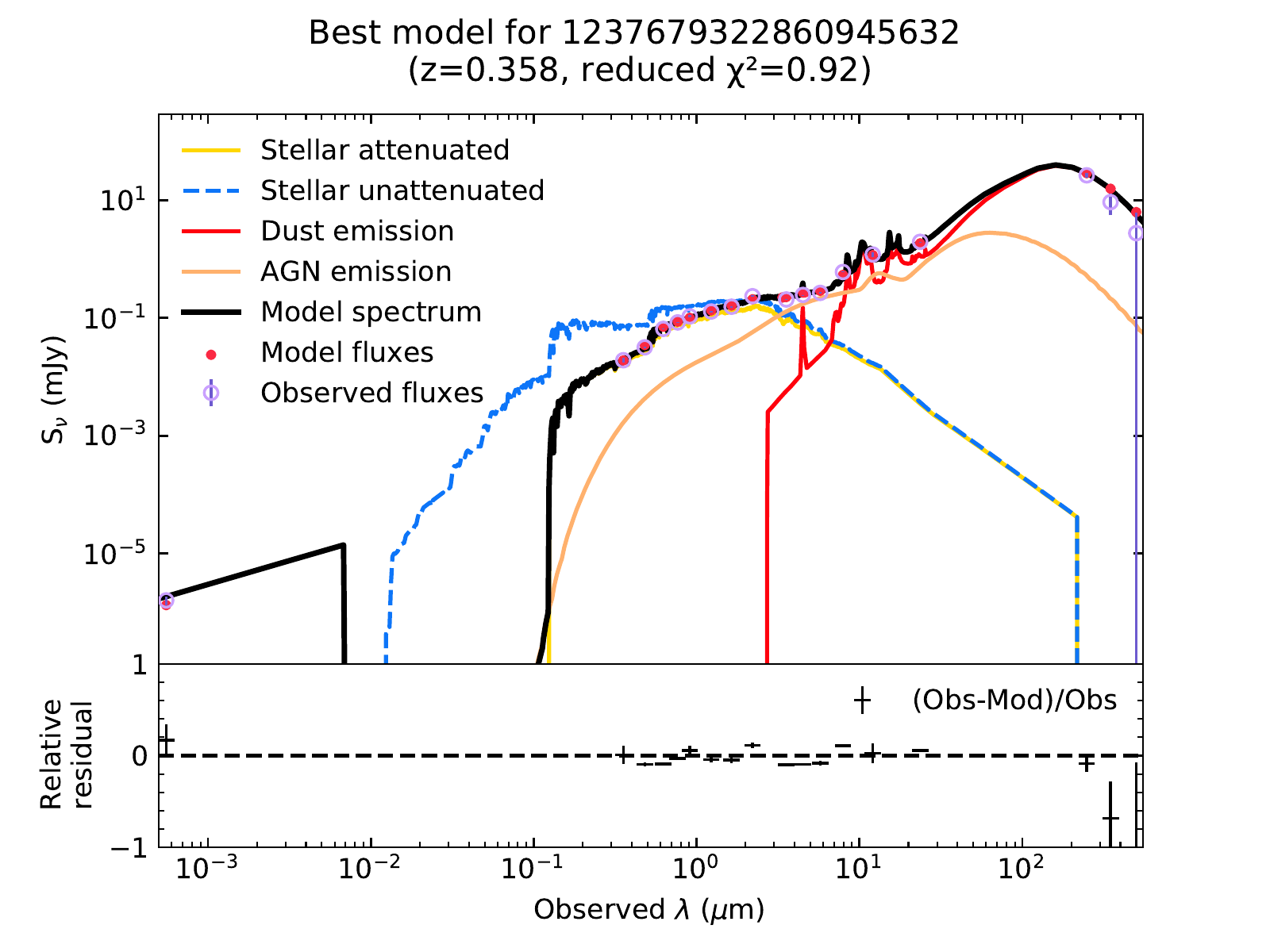}
   \label{} 
\end{subfigure}

\begin{subfigure}[b]{0.5\textwidth}
   \includegraphics[width=1\linewidth, height=7.2cm]{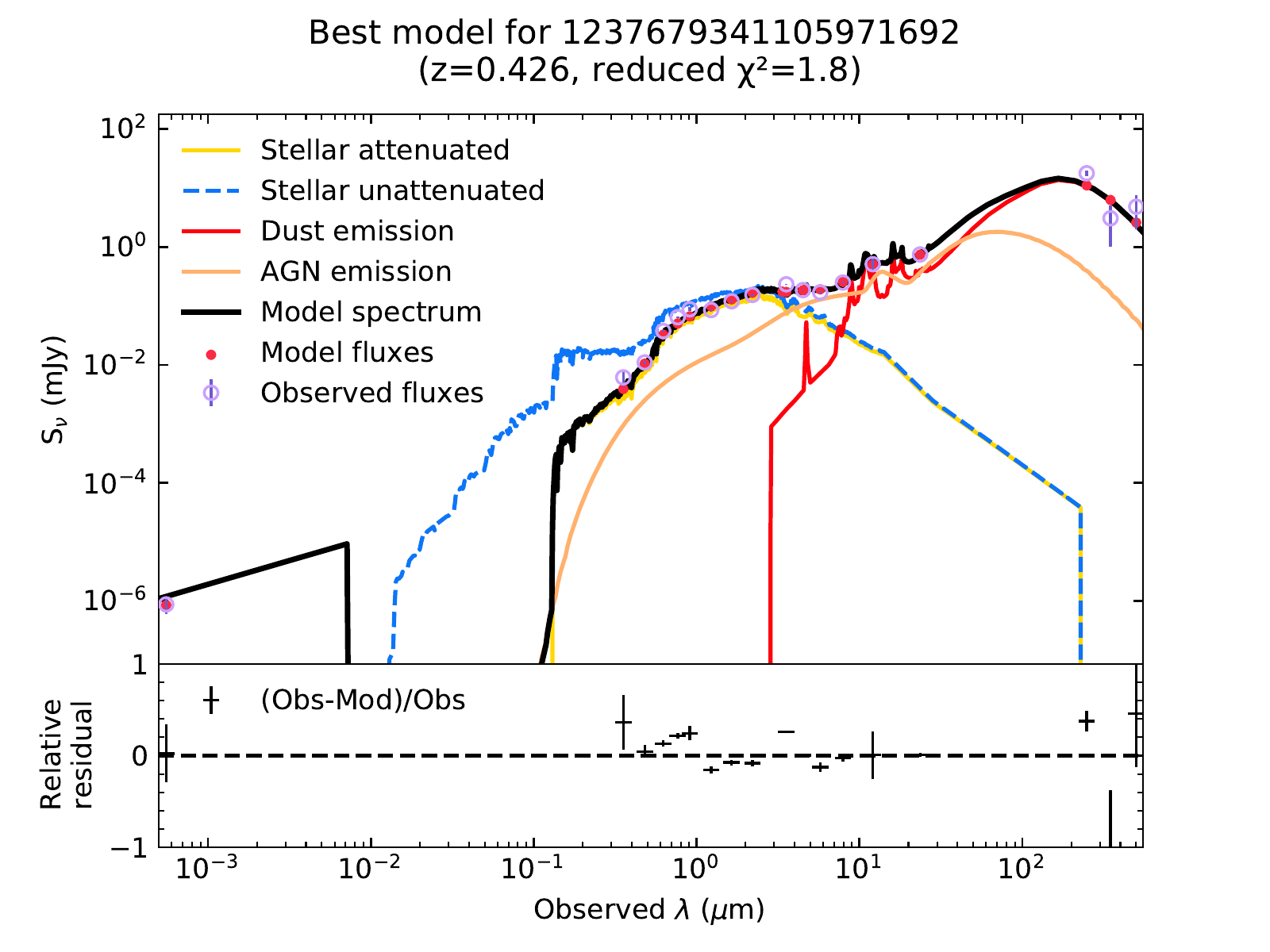}
   \label{}
\end{subfigure}

\begin{subfigure}[b]{0.5\textwidth}
   \includegraphics[width=1\linewidth, height=7.2cm]{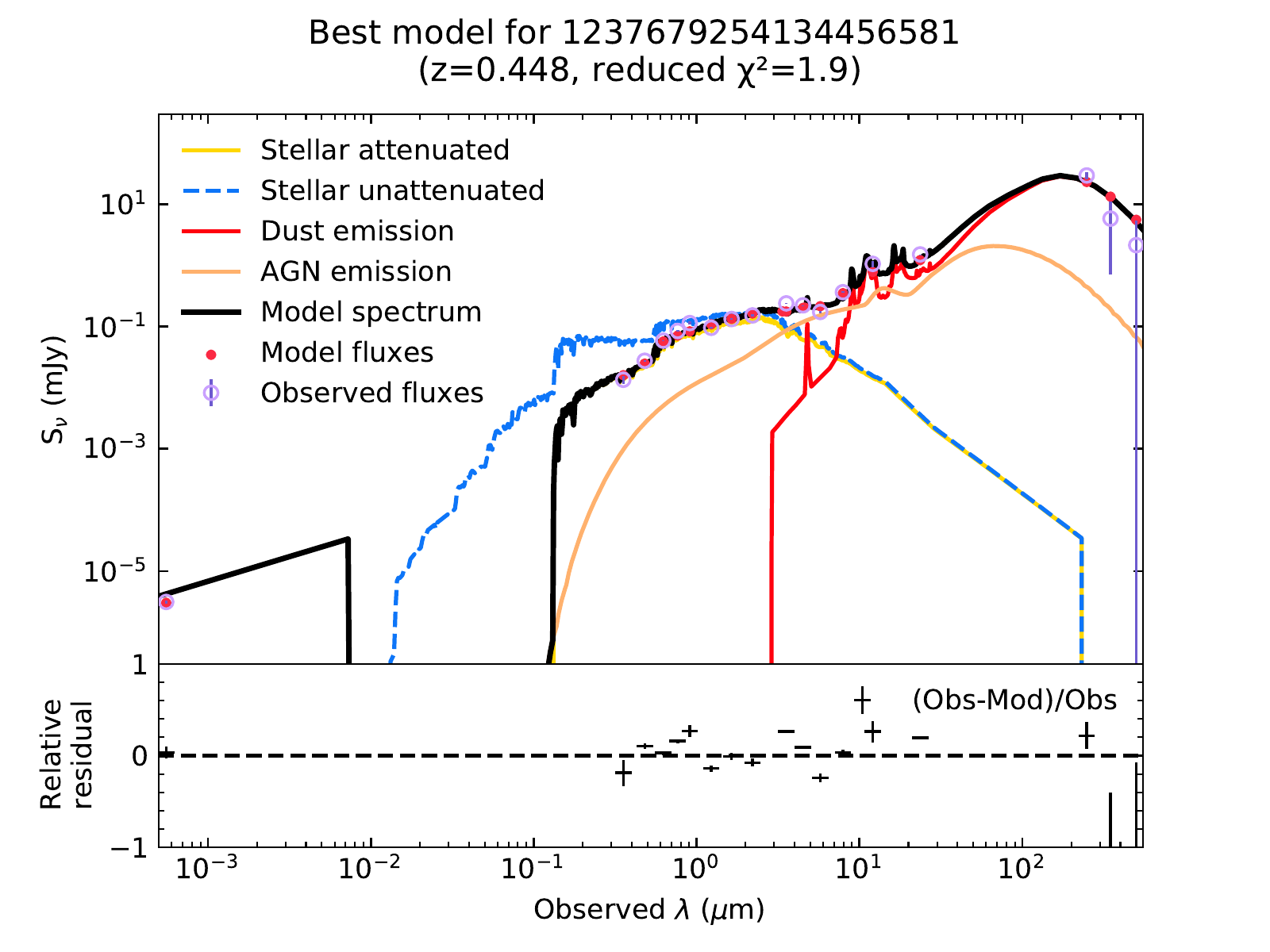}
   \label{}
\end{subfigure}

\caption{Examples of SEDs of the 27 type 2 sources that are classified as unobscured, based on inclination value estimated by X-CIGALE. We notice that their AGN emission presents (some) absorption in the optical part of the spectrum in agreement with their spectral classification.}
\label{fig_SEDs_type2_unobscured}
\end{figure}

\begin{figure}
\centering
   \includegraphics[width=1\linewidth, height=7.2cm]{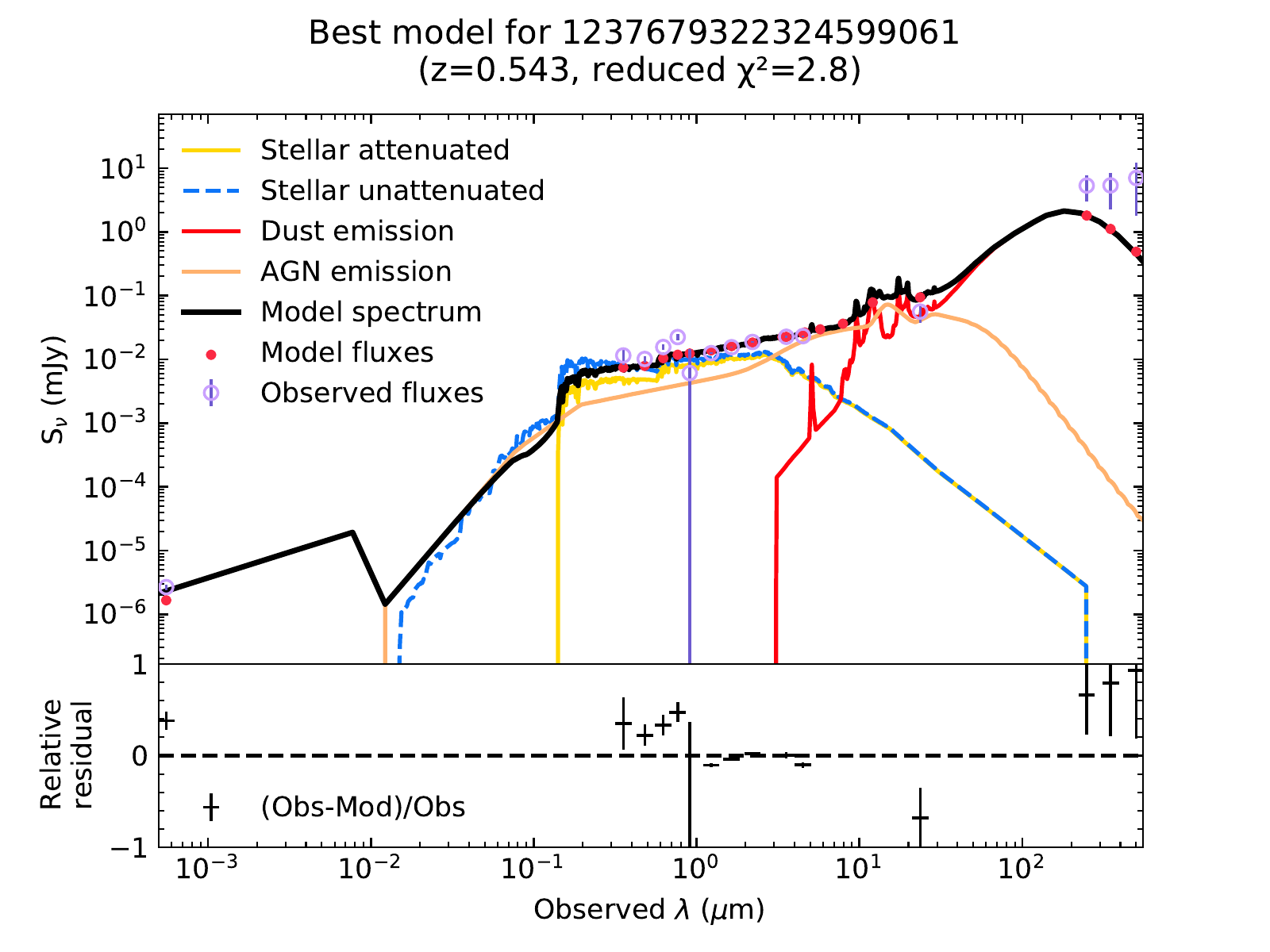}
\caption{SED of AGN that is spectroscopic type 2, but type 1 based on X-CIGALE and does not have significant polar dust ($\rm E_{(B-V), bayes}<0.15$). Although, the $\chi ^2_{red}$ is lower than the threshold we set to exclude sources, the optical and far-IR photometry is inconsistent and could not be fit by X-CIGALE.}
\label{fig_SEDs_type2_unobscured_problematic}
\end{figure}

\begin{figure}
\centering
\begin{subfigure}[b]{0.5\textwidth}
   \includegraphics[width=1\linewidth, height=7.2cm]{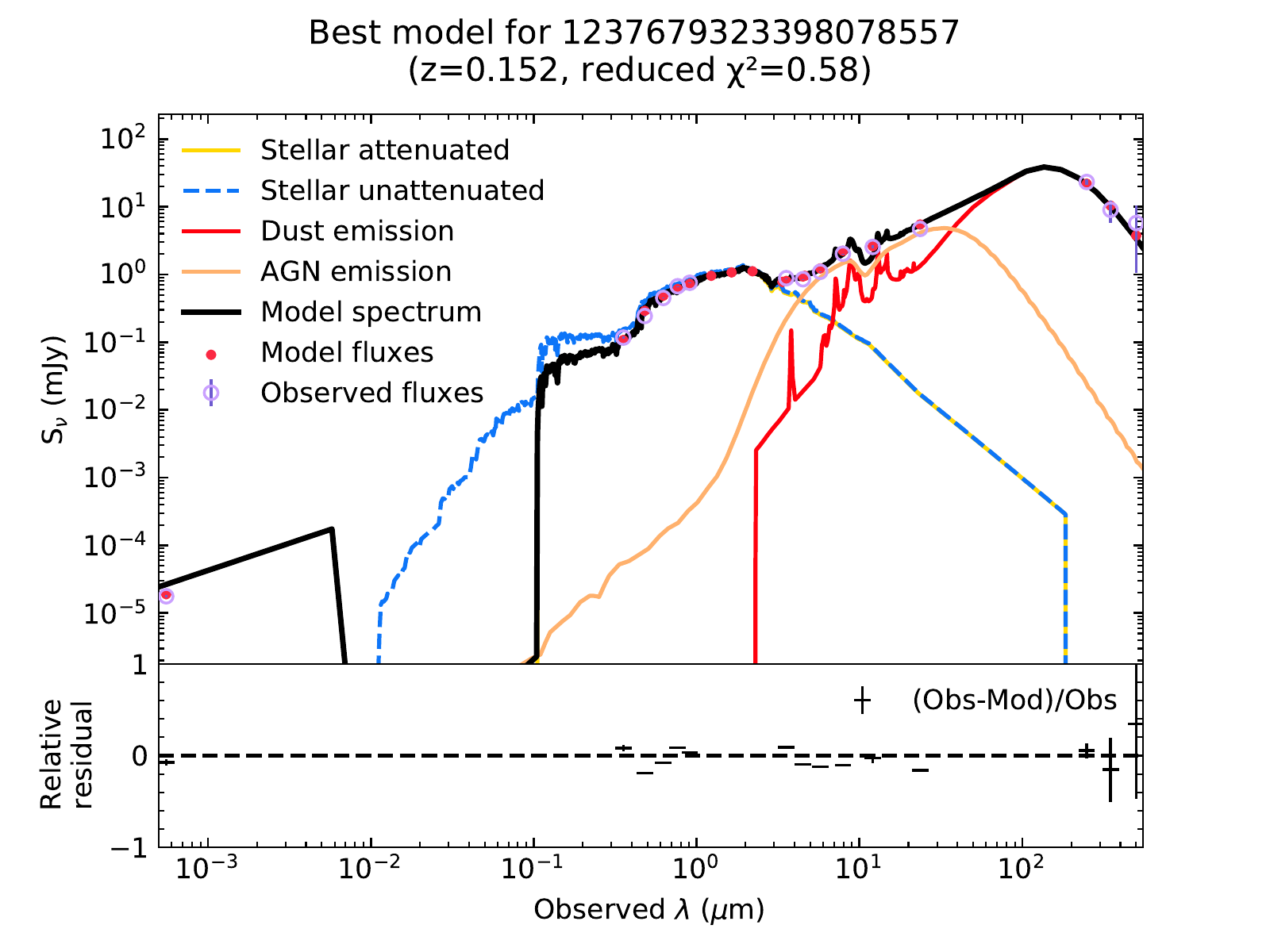}
   \label{} 
\end{subfigure}

\begin{subfigure}[b]{0.5\textwidth}
   \includegraphics[width=1\linewidth, height=7.2cm]{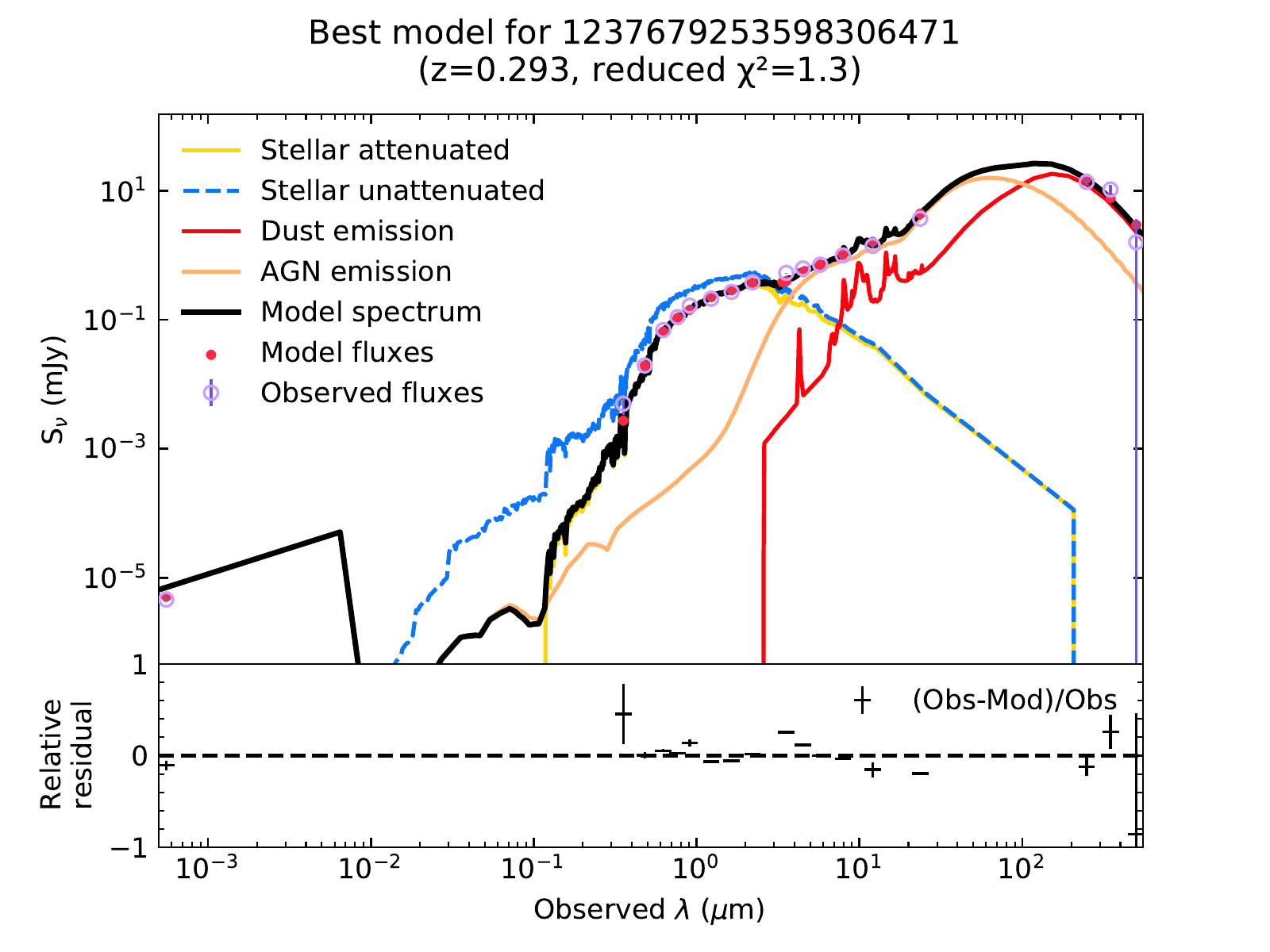}
   \label{}
\end{subfigure}

\begin{subfigure}[b]{0.5\textwidth}
   \includegraphics[width=1\linewidth, height=7.2cm]{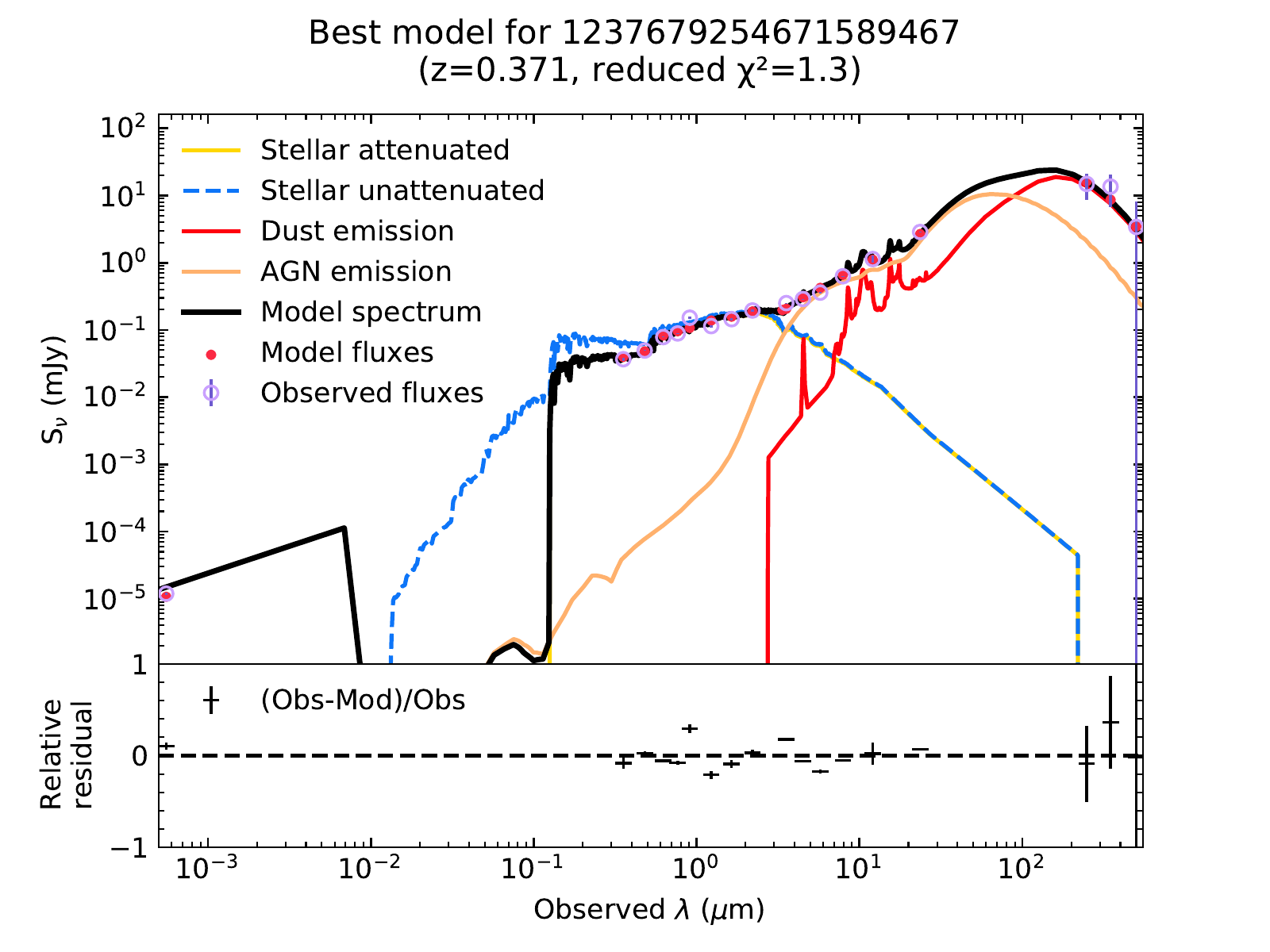}
   \label{}
\end{subfigure}

\caption{Examples of SEDs of the 28 spectroscopic type 1 sources that are classified as type 2, i.e., dust obscured with a viewing angle of $i=70^{\circ}$, based on X-CIGALE.}
\label{fig_SEDs_type1_obscured}
\end{figure}

\begin{figure*}
\centering
\begin{subfigure}{.5\textwidth}
  \centering
  \includegraphics[width=1\linewidth, height=7.2cm]{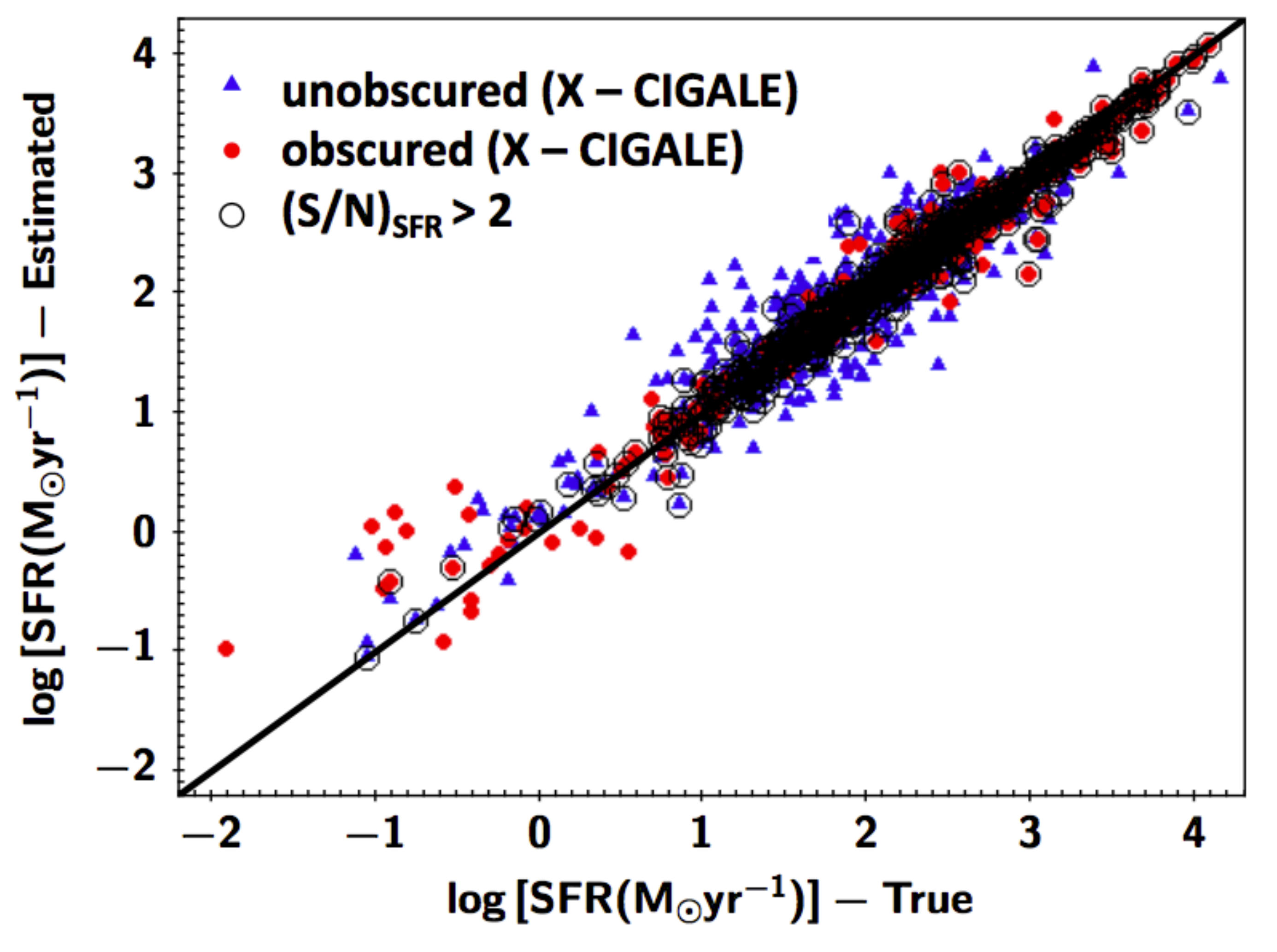}
  \label{}
\end{subfigure}%
\begin{subfigure}{.5\textwidth}
  \centering
  \includegraphics[width=1\linewidth, height=7.2cm]{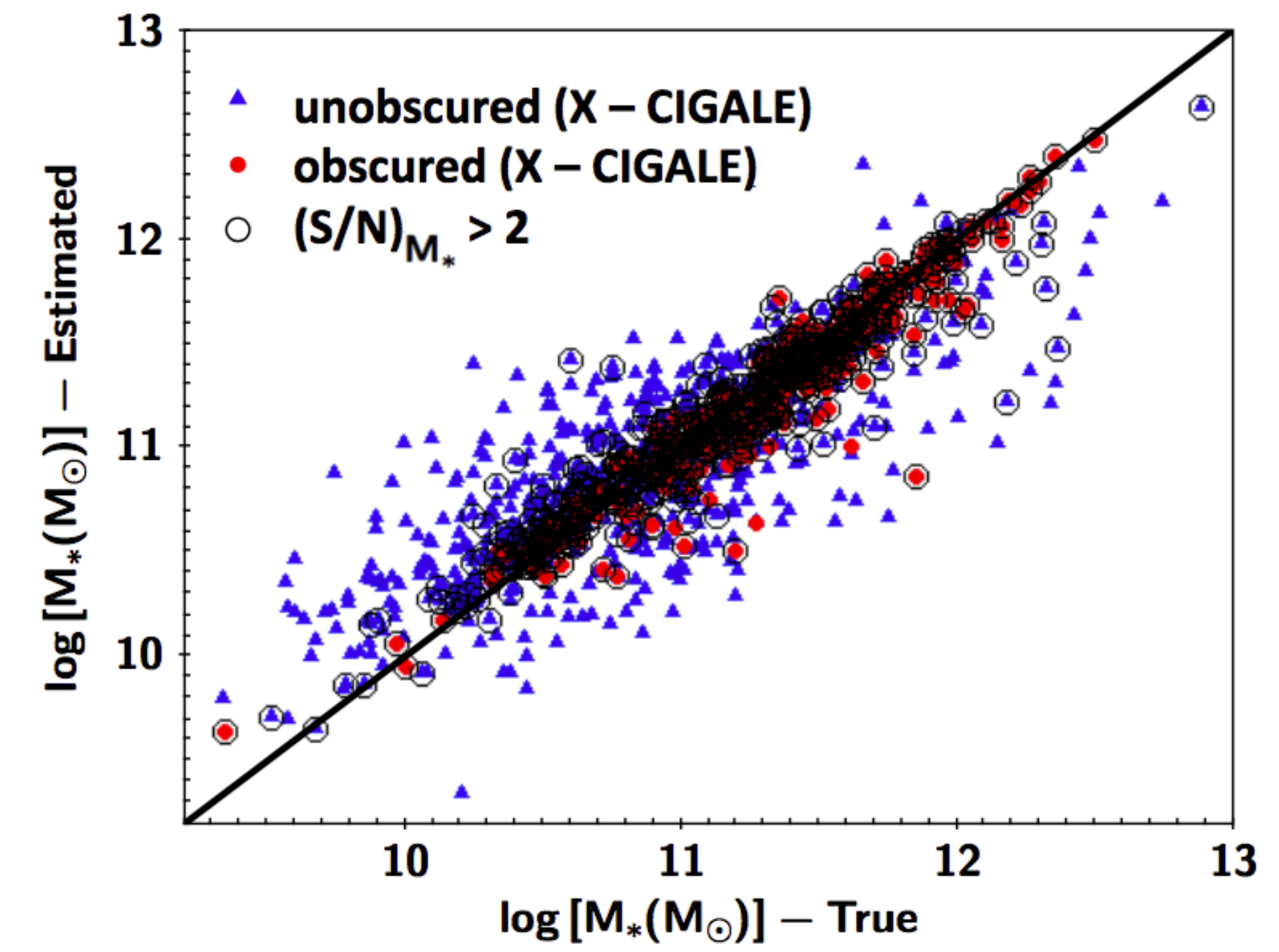}
  \label{}
\end{subfigure}
\caption{Comparison of the SFR and M$_*$ measurements (left and right panel, respectively) for the estimated and true values from the mock analysis. Blue triangles show the results for type 1 X-ray AGN and red circles for type 2. Sources are classified using the value of inclination angle, estimated by SED fitting. The black solid line shows the 1:1 relation. Restricting the measurements to those sources with statistical significance, $S/N>2$ (open circles), effectively reduces the scatter of the calculations.}
\label{fig_data_mock_cigale}
\end{figure*}

\begin{table*}
\caption{Comparison of classification of X-ray AGN, based on the inclination angle estimated by X-CIGALE and optical /\ mid-IR colours \citep{Hickox2017} with respect to the classification using optical spectroscopy. }
\centering
\begin{tabular}{ccccc}
& \multicolumn{2}{c}{type 1} & \multicolumn{2}{c}{type 2} \\
\hline
criterion & reliability & completeness & reliability & completeness \\
\hline
X-CIGALE & 86\% & 85\%& 47\% & 48\%  \\
X-CIGALE (E$_{B-V}=0$) & 95\% & 61\% & 42\% & 75\%\\
Hickox et al. & 92\%  & 50\% & 32\% & 85\% \\
\label{table_classification}
\end{tabular}
\tablefoot{Completeness refers to how many sources classified as type 1 (or 2) based on optical spectroscopy were identified as such by the other two classification criteria. The reliability is defined as the fraction of the number of type 1 (or 2) sources classified by SED fitting or colour criteria that are classified similarly by optical spectra. X-CIGALE completeness for identifying obscured systems rises to $100\%$, if we consider as type 2 systems that are type 1 based on their inclination angle but have increased polar polar dust. We also quote the percentages when we run X-CIGALE without the ability to add polar dust (E$_{B-V}=0$).}
\end{table*}


\section{Comparison of AGN classification using different criteria}
In the second part of our analysis, we compare the AGN classification based on optical spectra with that from SED fitting. Our goal is to investigate the consistency of the classification of X-CIGALE compared to optical spectra and examine whether the performance of the SED fitting classification affects the accuracy of the measurements of the host galaxy properties. We also examine the efficiency and effectiveness of optical/mid-IR colour compared to optical spectra and X-CIGALE source classification.

\subsection{Classification of AGN based on X-CIGALE}

\subsubsection{X-CIGALE vs. spectral classification and the effect of polar dust} 
\label{sec_xcigale_classif}

One of the parameters of the AGN module estimated via SED fitting is the viewing angle, $i$, at which the source is observed. We split X-ray AGN into two types based on the value of the inclination angle and compare their X-CIGALE classification with that from optical spectra. We restrict the AGN sample to the 284 AGN that have secure spectral classification and are within the same redshift and L$_X$ range (see Section \ref{sec_optspectra}). Examination of the mock catalogue reveals that X-CIGALE retrieves the $i$ parameter within $\pm 2^{\circ}$ (mean value of the difference of the estimated to the true $i$ values). The dispersion is $10^{\circ}$. In this exercise, we only consider AGN with a secure X-CIGALE  classification. To identify these sources, we use the bayes and best estimates of the $i$ parameter, derived by the SED fitting. Secure type 1 sources, based on X-CIGALE, are those with $i_{best}=30^{\circ}$ and $i_{bayes}<40^{\circ}$, while secure type 2 sources are those with $i_{best}=70^{\circ}$ and $i_{bayes}>60^{\circ}$. 240 ($\sim 85\%$) out of the 284 AGN, have secure classification from X-CIGALE. 187 are type 1 and 53 are type 2 AGN. The reasons that 44 sources do not have a secure classification are investigated in the Appendix. 

We also examine the classification performance of X-CIGALE for sources that lie at $\rm z>1$. Since at higher redshifts there are no spectroscopic type 2 AGN, we compare X-CIGALE's classification with that from optical spectra only for type 1 sources. Our analysis shows that X-CIGALE classifies type 1 AGN with similar efficiency  both at $\rm z<1$ and $\rm z>1$, when sufficient photometric coverage is available. The details of this analysis are presented in the Appendix.

Fig. \ref{fig_cigale_Lx_redz} presents the redshift, L$_X$ and N$_H$ distributions of the two AGN populations for the 284 sources. For comparison, we plot the same distributions for type 1 and 2 sources, based on optical spectra (dashed lines). The redshift and L$_X$ distributions of sources classified as type 1, based on X-CIGALE and optical spectra, are similar, peaking at higher redshifts and X-ray luminosities compared to their type 2 counterparts. The L$_X$ and redshift distributions of type 2 sources, based on SED fitting, appear flatter compared to spectroscopic type 2 AGN. The N$_H$ distribution of type 1 sources peaks at low N$_H$ values ($N_H<10^{21}\,\rm cm^{-2}$). The N$_H$ distribution of type 2 sources show a second peak at $N_H>10^{22}\,\rm cm^{-2}$. similar to the N$_H$ distribution of type 2 sources, classified based on optical spectra (right panel of Fig. \ref{fig_type_Lx_redz}). As previously noted, broad line AGN that are obscured in X-rays have been reported in previous studies \citep[e.g.][]{Merloni2014}.

Next, we compare the X-CIGALE classification with that from optical spectra, for the sample of 240 X-ray AGN. In the following, completeness refers to how many sources classified as type 1 (or type 2) based on optical spectroscopy were identified as such by the SED fitting results. The reliability is defined as the fraction of the number of type 1 (or type 2) sources classified by SED fitting that are classified similarly by optical spectra.

Fig. \ref{fig_comp_referee} presents the comparison of the two criteria. SED fitting efficiently recovers the majority (160/188=85\%, Table \ref{table_classification}) of spectroscopic type 1 AGN. Moreover, most of the sources (160/187=86\%) classified as type 1, based on the inclination angle estimates of X-CIGALE, are also identified similarly using optical spectra. The performance of SED fitting drops for type 2 sources. The reliability and completeness of X-CIGALE to uncover type 2 sources is 47\% (25/53) and 48\% (25/52), respectively (Table \ref{table_classification}).

The ability of X-CIGALE to model and quantify the presence of polar dust in AGN, was shown to improve e.g., the AGN fraction estimates of the SED fitting \citep{Mountrichas2021}. However, it makes more complex the definition of obscured and unobscured sources. Although polar dust does not affect the UV/optical SED of type 2 AGN, which is already absorbed by the dusty torus, it reddens the UV/optical SED of type 1 AGN. Thus, the polar dust model provides a physical explanation for red, type 1 AGN \citep{Yang2020}. Here, we examine how the addition of polar dust in the SED fitting process affects the comparison between X-CIGALE and spectral classification. First, we run X-CIGALE without the ability of adding polar dust (E$_{B-V}=0$) for all 240 AGN and compare its classification with that of optical spectra. In this case, X-CIGALE securely classifies 214/240 AGN. Setting E$_{B-V}=0.0$, lowers the efficiency of X-CIGALE to identify spectroscopic type 1 sources to $\sim 61\%$ (from $85\%$ when polar dust is considered) and increases it efficiency in recovering spectroscopic type 2 sources to $75\%$ (from $48\%$ when polar dust is considered). The percentages are shown in Table \ref{table_classification}. We note, that excluding the X-ray flux from the SED fitting process does not affect the classification of the sources.

There are 131 out of the 187 type 1 sources based on the inclination angle that have $\rm E_{(B-V), bayes}>0.15$ (105/131 are also classified as type 1, based on optical spectra). This value corresponds to substantial extinction in the UV ($\sim 50\%$), given the SMC extinction curve assumed for the SED modelling. For these sources, we examine whether the fits with polar dust are statistically different from those without polar dust. For that, we compute the Bayesian Information criterion (BIC). The two fits are then compared using the $\Delta\rm BIC$ parameter \citep[e.g.][]{Ciesla2018, Buat2019, Pouliasis2020}. Only three sources show strong preference for polar dust, based on the $\Delta$BIC parameter ($\Delta\rm BIC>6.0$). 45 out of the 131 AGN are classified as secure type 2 sources, based on the inclination angle values, when polar dust is not considered. 15/45 are also spectroscopic type 2 (none has $\Delta\rm BIC>6.0$). We conclude that a significant fraction ($45/131\approx 35\%$) of type 1 (X-CIGALE) systems with increased polar dust, could be classified securely as type 2 if polar dust is ignored, with the two classifications to not differ statistically. Thus, in the following, when we examine sources that have different classification from X-CIGALE and optical spectra, we take into account the effect of polar dust in these systems.


The comparison of the two classification criteria, presented in Fig. \ref{fig_comp_referee}, reveals that there are 27 sources that are type 2, based on optical spectra but classified as type 1, based on SED fitting. There are also 28 sources that are spectroscopic type 1 but type 2, based on X-CIGALE. We examine these sources further to understand possible reasons for their different classification.

Among the 27 sources that are spectroscopic type 2, but type 1 based on the inclination angle of X-CIGALE, 26 have $\rm E_{(B-V), bayes}>0.15$ and 15 are classified as type 2, when we run X-CIGALE without the option of polar dust (E$_{B-V}=0.0$). Visual inspection of the SEDs of these 26 AGN, showed that although they are classified as type 1 based on the inclination angle, their AGN emission presents (some) obscuration in the optical part of the spectrum. Fig. \ref{fig_SEDs_type2_unobscured}, presents three examples of these cases. The SED of the one source out of the 27, that is type 1 and with $\rm E_{(B-V), bayes}<0.15$, is presented in Fig. \ref{fig_SEDs_type2_unobscured_problematic}. We notice, that although the $\chi ^2_{red}$ is lower than the threshold we set to exclude sources, optical photometry appears problematic, possibly due to blending with nearby, bright optical sources, and thus the fit from X-CIGALE is unreliable.

Regarding the 28 X-ray AGN that are spectroscopic type 1, but type 2 based on X-CIGALE, 23 of them are systems with increased polar dust ($\rm E_{(B-V), bayes}>0.15$). High values of polar dust in type 2 systems imply that there is a strong mid-IR emission detected in these systems. We run X-CIGALE forcing the 28 sources to be type 1 ($i=30^{\circ}$). Based on $\Delta$BIC, only two sources strongly favour ($\Delta\rm BIC>6.0$) the fit with the run that has the inclination angle free. However, when the classification is forced to type 1 to match the spectral classification, polar dust, in most systems, increases further. Specifically, the mean  $\rm E_{(B-V), bayes}=0.28$ and $\rm E_{(B-V), bayes}=0.23$, when $i=30^{\circ}$ and $i$ is free, respectively. Examples of the SEDs of these 28 AGN are presented in Fig. \ref{fig_SEDs_type1_obscured}. Although these sources appear as type 1 based on the optical spectra, SED fitting analysis strongly suggests that these are absorbed systems. \cite{Merloni2014} found that a fraction of high luminosity AGN, present broad lines in their optical spectra, but have absorbed X-ray spectra. The mean N$_H$ of the 28 sources is increased compared to that of the 160 AGN that are classified as type 1, based on both optical spectra and SED fitting (21.2\,cm$^{-2}$ vs. 20.8\,cm$^{-2}$), but only 7/28 have N$_H>21.5$\,cm$^{-2}$. Previous studies have also reported similar cases of broad line X-ray AGN classified as type 2 based on SED fitting, without being necessarily X-ray absorbed \citep[e.g.,][see their Table 7 and Figures in their Appendix]{Masoura2020}. Different scenarios that allow a complex distribution of gas and dust in AGN have been suggested to explain the large variety of AGN properties \citep[e.g.][]{Lyu2018, Ogawa2021, Arredondo2021}. 


Overall, regarding the type 2 population of X-ray AGN, X-CIGALE identifies all of them, either as type 2 based on the inclination angle or as type 1 systems with increased polar dust. The only spectroscopically classified as type 2 source that has neither of the above, presents problematic optical photometry (Fig. \ref{fig_SEDs_type2_unobscured_problematic}). About the type 1 population of X-ray AGN, X-CIGALE identifies as type 1 160/188 sources. The 28 type 1 AGN that X-CIGALE classifies as type 2, could be systems with an extended, clumpy dust component along the polar direction.

\subsubsection{The effect of X-CIGALE classification on the accuracy of host galaxy property measurements}

In Section \ref{sec_optspectra}, we examined the reliability of the SFR and M$_*$ measurements of the SED fitting, when the inclination angle of each source is fixed to a value based on the classification from optical spectra. Now, we use the 1201 X-ray sources in our dataset (see Section \ref{sec_unreliable}) and repeat the same exercise, setting the viewing angle free to examine whether the misclassification of X-CIGALE for some AGN, affects the reliability of the calculations of the host galaxy properties. There are 972 ($\sim 81\%$) sources with secure classification from X-CIGALE. 681 are type 1 and 291 are type 2, based on SED fitting. We use the mock catalogues created by X-CIGALE and follow the procedure described in Section \ref{sec_optspectra}. Fig. \ref{fig_data_mock_cigale} presents the results. Sources that have $S/N>2$ are presented by open circles. The number of X-ray AGN that satisfy the aforementioned criterion are shown in Table \ref{table_numbers}. We conclude, that although in this case the classification of AGN was not fixed and thus some sources are misclassified by X-CIGALE, this does not affect the reliability of the host galaxy measurements. 



\begin{figure*}
\centering
  \centering
  \includegraphics[width=1\linewidth]{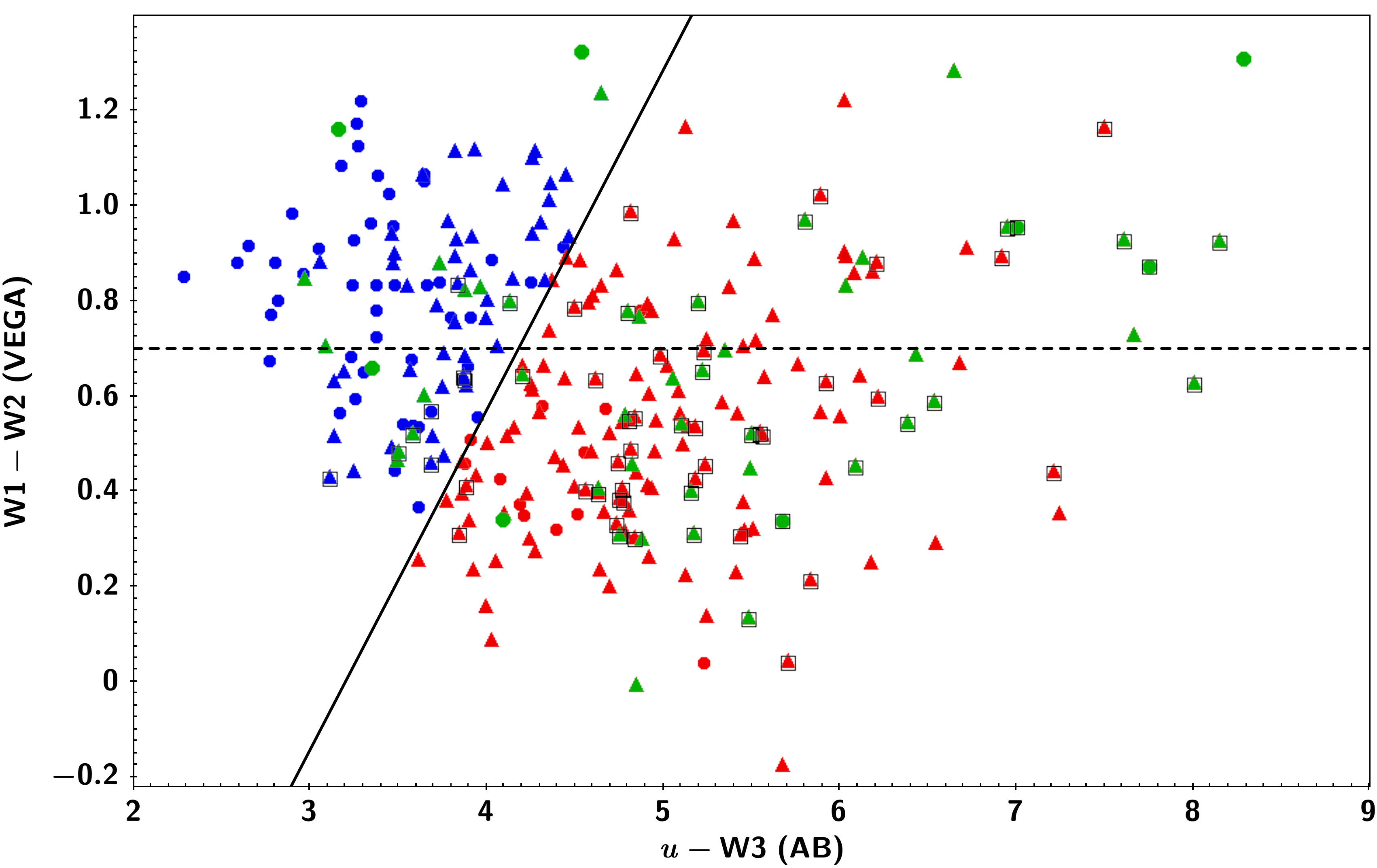}
  \label{}
\caption{The solid line splits the colour-colour diagram into red and non red AGN, using the Hickox et al. 2017 criterion. The horizontal dashed line indicates the $W1-W2>0.7$ limit for the selection of IR AGN. Sources classified as optical red, are shown in red, while the rest are shown in blue. AGN classified as type 2 from the SED fitting are presented in green. Sources with E$_{B-V}>0.15$ are shown with triangles. Sources spectroscopically classified as type 2, are marked by open squares.}
\label{fig_sed_vs_colour}
\end{figure*}

\subsection{Effectiveness of optical /\ mid-IR colours in classifying X-ray AGN}

In this Section, we discuss the performance of optical/mid-IR colour criteria to detect obscured (type 2) AGN. The obscured AGN population is known to present very red optical/mid-IR colours, due to the extinction of the nuclear emission in (rest-frame) optical and UV wavelengths \citep[e.g.][]{Hickox2007}. Although these colour criteria are very efficient in uncovering obscured AGN selected in the mid-IR, X-rays are known to miss a large fraction of obscured sources selected based on optical/mid-IR colours. \cite{Mountrichas2020} used sources detected in Stripe 82 and found that among IR selected AGN with SDSS detection, $43\%$ are optically red. However, this percentage drops to $23\%$ among those AGN that are also X-ray detected. \cite{Masoura2020}, used data from the XMM-XXL and found that only $\sim25\%$ of red, IR selected AGN are detected in X-rays. Therefore, our X-ray selected AGN sample is biased against optical red sources.

Different optical/mid-IR criteria exist in the literature that classify sources into obscured and unobscured. \cite{Yan2013}, proposed that red sources are identified using $\rm {\it{r}}-W2>6$. However, this criterion is not sensitive at redshifts below $\rm z<0.5$ \citep{Yan2013, Hickox2017}. This is also true for other similar criteria \citep[e.g.,][]{Hickox2007, Lamassa2016}. Thus, we choose to apply an optical/mid-IR criterion that effectively separates AGN into type 1 and 2 at all redshifts and most importantly at low redshift ($\rm z<1$) where our spectroscopic type 2 sources lie. Thus, we select red AGN among the 240 X-ray sources (see Section \ref{sec_xcigale_classif}), by applying the criterion presented in \cite{Hickox2017}. Specifically, red sources are identified using $\rm {\it{u}}-W3\,[AB]>1.4(W1-W2\,[VEGA])+3.2$. 



The comparison of \cite{Hickox2017} optical/mid-IR criteria with optical spectra is shown in Fig. \ref{fig_comp_referee} and Table \ref{table_classification}. The colour criterion classifies many sources as red ($138/240\approx 57\%$). It recovers most of spectroscopic type 2 AGN, but also includes a large number of spectroscopic type 1 AGN. Only 32\% ($44/138$) of red AGN are spectroscopically classified as type 2. Moreover, only half of type 1 sources are identified as non red systems ($94/188$). \cite{Hickox2017} find that the criterion identifies $>90\%$ of spectroscopic type 1 and 2 AGN. We note that their sample consists of SDSS (luminous) quasars. Our sample, although consists of sources observed by SDSS, is X-ray selected and therefore includes sources with lower and moderate luminosities in addition to luminous AGN. We select the most luminous sources in our sample, by applying the following criteria: $\rm W1-W2>0.7$ and $\rm W2<15.05$ \citep[e.g., ][]{Stern2012}. There are 11 and 67 spectroscopic type 2 and type 1 sources that satisfy these criteria. Hickox et al. criterion identifies 28 red and 50 non red sources. The first noticeable difference is that in this more luminous AGN subsample, the fraction of red sources is significantly lower ($28/78=36\%$). This is expected since the fraction of obscured sources drops as the luminosity increases. 8 sources are both red and type 2, i.e. $\sim 73\%$ of spectroscopy type 2 are also red, while 47/67 ($\sim 70\%$) are type 1 and non red. These numbers are closer to the percentages of correct identifications quoted in \cite{Hickox2017}. This indicates that optical/mid-IR colours could be less effective in separating X-ray AGN into obscured and unobscured, compared to optically selected quasars and IR selected AGN. IR and optical selected AGN samples are dominated by luminous AGN (e.g. $\rm log\,[L_{X}(ergs^{-1})]>44$) and a clear separation is observed between unobscured AGN, that are optically bright, and obscured AGN, for which we only observe their stellar emission. On the other hand, X-ray detected AGN include a large fraction of moderate to low luminosity AGN that blur the separation of optical magnitude distributions of the two AGN types \citep{Georgakakis2020}. 


Figure \ref{fig_sed_vs_colour}, shows the colour-colour space diagram, that Hickox et al. used to define their criterion (shown by the solid line). The plot is made using magnitudes from the data, i.e., not those estimated by X-CIGALE during the SED fitting. Sources classified as red are shown in red colour and lie on the right side of the solid line. AGN classified as type 2 from the SED fitting are shown in green, while those spectroscopically classified as type 2 are marked with open squares. Sources with E$_{B-V}>0.15$ are presented with triangles. Among the 138 sources classified as red, by the Hickox et al. criterion, only 39 are type 2, based on X-CIGALE. However, most red sources are systems with increased amount of polar dust (121/138=$88\%$), because of the strong reddening of their optical continuum. In the remaining 99 AGN (138-39), only 22 are spectroscopically classified as type 2. We note that all these 22 sources have  E$_{B-V}>0.15$, i.e. are AGN that although are classified as type 1 by X-CIGALE, based on their inclination angle, they have increased polar dust. This can be, at least partially, attributed to the fact that different obscuration criteria are sensitive to different levels of obscuration \citep[e.g.,][]{Mountrichas2019, Masoura2020}. Finally, among the 28 sources that are spectroscopically classified as type 1 but as type 2, based on X-CIGALE (see previous Section), 17 are red systems based on the colour criterion.



Our analysis is based on SED fitting of individual AGN which increases the diversity of the examined cases. Inclusion of different stellar spectra and the ability to model the AGN absorption (polar dust) are some of the factors that increase the complexity and introduce degeneracies that could blur the boundaries of the locus that the two AGN populations occupy in the optical/mid-IR colour space. The effect of polar dust on the AGN classification also depends on the adopted extinction curve and could change significantly if a flatter extinction curve \citep[e.g.,][]{Gaskell2004} is adopted instead of the SMC that is used in our analysis. Moreover, SED fitting requires good quality of photometric data and large wavelength coverage. In cases that these criteria are not satisfactory met, the errors on the calculated fluxes can be quite large and the final fits do not necessarily reproduce well the colours. The effects of these factors will be investigated in an upcoming paper.

\section{Summary}

In this work, we study a sample of spectroscopic, X-ray selected AGN in the XMM-XXL field, which have type 1 and type 2 classifications based on their optical spectra from \cite{Menzel2016}. Our goal is to examine the host galaxy properties of the two AGN populations and compare the spectroscopic classification of AGN with that from SED fitting. 

To estimate the SFR and M$_*$, we construct SEDs for 1,577 X-ray AGN, using optical to far-IR photometry. About half of the sources have been observed by {\it{Herschel}}. The SEDs are fit using the X-CIGALE code. X-CIGALE allows the inclusion of the X-ray flux in the fitting process and has the ability to account for extinction of the UV and optical emission in the poles of AGN, by modelling polar dust. We restrict our analysis to those X-ray sources that meet our photometric criteria for available optical and mid-IR photometry, have secure optical classification and reliable estimates from the SED fitting.

The redshift and X-ray luminosities of spectroscopic type 1 and type 2 AGN, present very different distributions (Fig. \ref{fig_type_Lx_redz}). To compare their host galaxy properties, we match their L$_X$ and redshift distributions, by weighting each source. This effectively reduces our sample to 284 sources. Our analysis reveals that both AGN populations live in galaxies with similar SFR and SFR$_{norm}$. The latter is defined as the ratio of the SFR of AGN to the SFR of star-forming main sequence (MS) galaxies with the same stellar mass and redshift. However, type 2 AGN tend to reside in more massive hosts compared to their type 1 counterparts. Specifically the average stellar mass of type 1 host galaxies is $10.57^{+0.20}_{-0.12}\,\rm M_\odot$, compared to $10.87^{+0.06}_{-0.12}\,\rm M_\odot$ for type 2. Although, this result is statistical significant only at $\approx 1\,\sigma$, most likely due to the small examined sample, it is in agreement with previous studies \citep{Zou2019}.  

One of the parameters estimated by the SED fitting process is the inclination angle, $i$, that each AGN is observed. We compare the classification from X-CIGALE, based on the inclination angle, with that from optical spectra. X-CIGALE classifies as type 2 all spectroscopic type 2 sources, either by determining an edge on inclination angle or by measuring increased presence of polar dust in these systems. The algorithm, also successfully identifies the vast majority of type 1 sources ($160/188\approx 85\%$). There are 28 type 1 sources that X-CIGALE securely classifies as type 2. Seven of them are also X-ray obscured (N$_H>21.5$\,cm$^{-2}$). Visual inspection of their SEDs shows that these AGN experience either a strong absorption in the optical part of the spectrum and/or large contribution of polar dust. Similar sources have also been found in previous studies \citep[e.g.,][]{Merloni2014, LiuT2018, Masoura2020}. This class of AGN could be systems observed face on, that explains the presence of broad lines in their optical spectra with an extended dust component along the polar direction. This dust obscures the central SMBH causing an excess of mid-IR emission \citep{Lyu2018}. Similar results are found at $\rm z>1$, under the condition that sufficient and robust photometric data are available.

Finally, we compare the classification from SED fitting and optical spectra with that from optical/mid-IR colours using the criteria of \cite{Hickox2017}. $\sim 30\%$ of red sources are identified as type 2 based on X-CIGALE/spectra. However, this percentage increases to $\sim 75\%$, if we restrict the X-ray AGN sample to those sources that are also IR selected AGN ($W1-W2>0.7$). Therefore, optical/mid-IR colours seem to be more reliable in identifying obscured sources among IR selected AGN.


\begin{acknowledgements}
The authors thank the anonymous referee for their detailed report that improved the quality of the paper.
\\
GM acknowledges support by the Agencia Estatal de Investigación, Unidad de Excelencia María de Maeztu, ref. MDM-2017-0765.
\\
The project has received funding from Excellence Initiative of Aix-Marseille University - AMIDEX, a French `Investissements d'Avenir' programme.
\\
VAM and IG acknowledge support of this work by Greece and the European Union (European Social Fund-ESF) through the Operational Programme "Human Resources Development, Education and Lifelong Learning 2014-2020" in the context of the project "Anatomy of galaxies: their stellar and dust content though cosmic time" (MIS 5052455).
\\
MB acknowledges FONDECYT regular grant 1170618
\\
\\
XXL  is  an  international  project  based  around  an  $\it{XMM}$  Very Large Programme surveying two 25 deg$^2$ extragalactic fields at a depth of $\sim$  6 $\times$ $10^{-15}$ erg cm $^{-2}$ s$^{-1}$ in the [0.5-2] keV band for point-like sources. The XXL website ishttp://irfu.cea.fr/xxl/.  Multi-band  information  and  spectroscopic  follow-up  of the X-ray sources are obtained through a number of survey programmes,  summarised  at http://xxlmultiwave.pbworks.com/.
\\
This research has made use of data obtained from the 3XMM XMM-\textit{Newton} 
serendipitous source catalogue compiled by the 10 institutes of the XMM-\textit{Newton} 
Survey Science Centre selected by ESA.
\\
This work is based on observations made with XMM-\textit{Newton}, an ESA science 
mission with instruments and contributions directly funded by ESA Member States 
and NASA. 
\\
Funding for the Sloan Digital Sky Survey IV has been provided by the Alfred P. Sloan Foundation, the U.S. Department of Energy Office of Science, and the Participating Institutions. SDSS-IV acknowledges
support and resources from the Center for High-Performance Computing at
the University of Utah. The SDSS web site is \url{www.sdss.org}.
\\
SDSS-IV is managed by the Astrophysical Research Consortium for the 
Participating Institutions of the SDSS Collaboration including the 
Brazilian Participation Group, the Carnegie Institution for Science, 
Carnegie Mellon University, the Chilean Participation Group, the French Participation Group, Harvard-Smithsonian Center for Astrophysics, 
Instituto de Astrof\'isica de Canarias, The Johns Hopkins University, 
Kavli Institute for the Physics and Mathematics of the Universe (IPMU) / 
University of Tokyo, Lawrence Berkeley National Laboratory, 
Leibniz Institut f\"ur Astrophysik Potsdam (AIP),  
Max-Planck-Institut f\"ur Astronomie (MPIA Heidelberg), 
Max-Planck-Institut f\"ur Astrophysik (MPA Garching), 
Max-Planck-Institut f\"ur Extraterrestrische Physik (MPE), 
National Astronomical Observatories of China, New Mexico State University, 
New York University, University of Notre Dame, 
Observat\'ario Nacional / MCTI, The Ohio State University, 
Pennsylvania State University, Shanghai Astronomical Observatory, 
United Kingdom Participation Group,
Universidad Nacional Aut\'onoma de M\'exico, University of Arizona, 
University of Colorado Boulder, University of Oxford, University of Portsmouth, 
University of Utah, University of Virginia, University of Washington, University of Wisconsin, 
Vanderbilt University, and Yale University.
\\
This publication makes use of data products from the Wide-field Infrared Survey 
Explorer, which is a joint project of the University of California, Los Angeles, 
and the Jet Propulsion Laboratory/California Institute of Technology, funded by 
the National Aeronautics and Space Administration. 
\\
The VISTA Data Flow System pipeline processing and science archive are described 
in \cite{Irwin2004}, \cite{Hambly2008} and \cite{Cross2012}. Based on 
observations obtained as part of the VISTA Hemisphere Survey, ESO Program, 
179.A-2010 (PI: McMahon). We have used data from the 3rd data release.
\\
This work is based [in part] on observations made with the Spitzer Space Telescope, which was operated by the Jet Propulsion Laboratory, California Institute of Technology under a contract with NASA
\\

\end{acknowledgements}

\bibliography{mybib}{}

\begin{thebibliography}{74}
\expandafter\ifx\csname natexlab\endcsname\relax\def\natexlab#1{#1}\fi

\bibitem[{Aird {et~al.}(2015)Aird, Coil, Georgakakis, Nandra, Barro, \&
  P{\'{e}}rez-Gonz{\'{a}}lez}]{Aird2015}
Aird, J., Coil, A.~L., Georgakakis, A., {et~al.} 2015, Monthly Notices of the
  Royal Astronomical Society, 451, 1892

\bibitem[{Antonucci(1993)}]{Antonucci1993}
Antonucci, R. 1993, Annual Review of Astronomy and Astrophysics, 31, 473

\bibitem[{Arredondo {et~al.}(2021)Arredondo, Martín, Dultzin, Masegosa,
  Almeida, Bernete, Fritz, \& Clavijo}]{Arredondo2021}
Arredondo, D.~E., Martín, O.~G., Dultzin, D., {et~al.} 2021
  [\eprint{http://arxiv.org/abs/2104.11263v1}]

\bibitem[{Ballantyne(2017)}]{Ballantyne2017b}
Ballantyne, D.~R. 2017, Monthly Notices of the Royal Astronomical Society, 464,
  626

\bibitem[{Bernhard {et~al.}(2019)Bernhard, Grimmett, Mullaney, Daddi,
  Tadhunter, \& Jin}]{Bernhard2019}
Bernhard, E., Grimmett, L.~P., Mullaney, J.~R., {et~al.} 2019, Monthly Notices
  of the Royal Astronomical Society: Letters, 483, L52

\bibitem[{Boquien {et~al.}(2019)Boquien, Burgarella, Roehlly, Buat, Ciesla,
  Corre, Inoue, \& Salas}]{Boquien2019}
Boquien, M., Burgarella, D., Roehlly, Y., {et~al.} 2019, Astronomy {\&}
  Astrophysics, 622, A103

\bibitem[{Bruzual \& Charlot(2003)}]{Bruzual_Charlot2003}
Bruzual, G. \& Charlot, S. 2003, MNRAS, 344, 1000

\bibitem[{Buat {et~al.}(2019)Buat, Ciesla, Boquien, Ma{\l}ek, \&
  Burgarella}]{Buat2019}
Buat, V., Ciesla, L., Boquien, M., Ma{\l}ek, K., \& Burgarella, D. 2019,
  Astronomy {\&} Astrophysics, 632, A79

\bibitem[{Buchner {et~al.}(2017)Buchner, Schulze, \& Bauer}]{Buchner2017}
Buchner, J., Schulze, S., \& Bauer, F.~E. 2017, Monthly Notices of the Royal
  Astronomical Society, 464, 4545

\bibitem[{{Buchner} {et~al.}(2014)}]{Buchner2014}
{Buchner}, J. {et~al.} 2014, A\&A, 564, 125

\bibitem[{Chabrier(2003)}]{Chabrier2003}
Chabrier, G. 2003, PASP, 115, 763

\bibitem[{Charlot \& Fall(2000)}]{Charlot_Fall_2000}
Charlot, S. \& Fall, S.~M. 2000, ApJ, 539, 718

\bibitem[{Chen {et~al.}(2015)Chen, Hickox, Alberts, Harrison, Alexander, Assef,
  Brodwin, Brown, Moro, Forman, Gorjian, Goulding, Hainline, Jones, Kochanek,
  Murray, Pope, Rovilos, \& Stern}]{Chen2015}
Chen, C.-T.~J., Hickox, R.~C., Alberts, S., {et~al.} 2015, The Astrophysical
  Journal, 802, 50

\bibitem[{Ciesla {et~al.}(2018)Ciesla, Elbaz, Schreiber, Daddi, \&
  Wang}]{Ciesla2018}
Ciesla, L., Elbaz, D., Schreiber, C., Daddi, E., \& Wang, T. 2018, Astronomy
  {\&} Astrophysics, 615, A61

\bibitem[{Ciotti \& Ostriker(1997)}]{Ciotti1997}
Ciotti, L. \& Ostriker, J.~P. 1997, The Astrophysical Journal, 487, L105

\bibitem[{Circosta {et~al.}(2019)Circosta, Vignali, Gilli, Feltre, Vito,
  Calura, Mainieri, Massardi, \& Norman}]{Circosta2019}
Circosta, C., Vignali, C., Gilli, R., {et~al.} 2019, Astronomy {\&}
  Astrophysics, 623, A172

\bibitem[{Cross {et~al.}(2012)}]{Cross2012}
Cross, N. J.~G. {et~al.} 2012, A\&A, 548, 21

\bibitem[{{Dale} {et~al.}(2014){Dale}, {Helou}, {Magdis}, {Armus},
  {D{\'{\i}}az-Santos}, \& {Shi}}]{Dale2014}
{Dale}, D.~A., {Helou}, G., {Magdis}, G.~E., {et~al.} 2014, ApJ, 784, 83

\bibitem[{Emerson {et~al.}(2006)Emerson, McPherson, \&
  Sutherland}]{Emerson2006}
Emerson, J., McPherson, A., \& Sutherland, W. 2006, Msngr, 126, 41

\bibitem[{Feltre {et~al.}(2012)Feltre, {Hatziminaoglou}, Fritz, \&
  {Franceschini}}]{Feltre2012}
Feltre, A., {Hatziminaoglou}, E., Fritz, J., \& {Franceschini}, A. 2012, MNRAS,
  426, 120

\bibitem[{Garcet {et~al.}(2007)Garcet, Gandhi, Gosset, Sprimont, Surdej,
  Borkowski, Tajer, Pacaud, Pierre, Chiappetti, Maccagni, Page, Carrera, Tedds,
  Mateos, Krumpe, Contini, Corral, Ebrero, Gavignaud, Schwope, F{\`{e}}vre,
  Polletta, Rosen, Lonsdale, Watson, Borczyk, \& Vaisanen}]{Garcet2007}
Garcet, O., Gandhi, P., Gosset, E., {et~al.} 2007, Astronomy {\&} Astrophysics,
  474, 473

\bibitem[{Gaskell {et~al.}(2004)Gaskell, Goosmann, Antonucci, \&
  Whysong}]{Gaskell2004}
Gaskell, C.~M., Goosmann, R.~W., Antonucci, R. R.~J., \& Whysong, D.~H. 2004,
  The Astrophysical Journal, 616, 147

\bibitem[{Georgakakis {et~al.}(2020)Georgakakis, Ruiz, \&
  LaMassa}]{Georgakakis2020}
Georgakakis, A., Ruiz, A., \& LaMassa, S.~M. 2020, Monthly Notices of the Royal
  Astronomical Society, 499, 710

\bibitem[{Hambly {et~al.}(2008)}]{Hambly2008}
Hambly, N.~C. {et~al.} 2008, MNRAS, 384, 637

\bibitem[{Hickox {et~al.}(2007)Hickox, Jones, Forman, Murray, Brodwin, Brown,
  Eisenhardt, Stern, Kochanek, Eisenstein, Cool, Jannuzi, Dey, Brand, Gorjian,
  \& Caldwell}]{Hickox2007}
Hickox, R.~C., Jones, C., Forman, W.~R., {et~al.} 2007, The Astrophysical
  Journal, 671, 1365

\bibitem[{Hickox {et~al.}(2017)Hickox, Myers, Greene, Hainline, Zakamska, \&
  DiPompeo}]{Hickox2017}
Hickox, R.~C., Myers, A.~D., Greene, J.~E., {et~al.} 2017, The Astrophysical
  Journal, 849, 53

\bibitem[{{Hickox} {et~al.}(2011)}]{Hickox2011}
{Hickox}, R.~C. {et~al.} 2011, ApJ, 731, 117

\bibitem[{Hopkins {et~al.}(2006)Hopkins, Hernquist, Cox, Matteo, Robertson, \&
  Springel}]{Hopkins2006}
Hopkins, P.~F., Hernquist, L., Cox, T.~J., {et~al.} 2006, The Astrophysical
  Journal Supplement Series, 163, 1

\bibitem[{Hönig {et~al.}(2006)Hönig, Beckert, Ohnaka, \&
  Weigelt}]{Hoenig2006}
Hönig, S.~F., Beckert, T., Ohnaka, K., \& Weigelt, G. 2006, Astronomy {\&}
  Astrophysics, 452, 459

\bibitem[{Irwin {et~al.}(2004)}]{Irwin2004}
Irwin, M.~J. {et~al.} 2004, SPIE, 5493, 411

\bibitem[{{Just} {et~al.}(2007){Just}, {Brandt}, {Shemmer}, {Steffen},
  {Schneider}, {Chartas}, \& {Garmire}}]{Just2007}
{Just}, D.~W., {Brandt}, W.~N., {Shemmer}, O., {et~al.} 2007, ApJ, 685, 1004

\bibitem[{{LaMassa} {et~al.}(2016)}]{Lamassa2016}
{LaMassa}, S.~M. {et~al.} 2016, ApJ, 818, 88

\bibitem[{{Lanzuisi} {et~al.}(2017)}]{Lanzuisi2017}
{Lanzuisi}, G. {et~al.} 2017, A\&A, 602, 13

\bibitem[{Li {et~al.}(2019)Li, Xue, Sun, Liu, Vito, Brandt, Hughes, Yang,
  Tozzi, Zhu, Zheng, Luo, Chen, Vignali, Gilli, \& Shu}]{Li2019}
Li, J., Xue, Y., Sun, M., {et~al.} 2019, The Astrophysical Journal, 877, 5

\bibitem[{Liu {et~al.}(2018)Liu, Merloni, Wang, Tozzi, Shen, Brusa, Salvato,
  Nandra, Comparat, Liu, Ponti, \& Coffey}]{LiuT2018}
Liu, T., Merloni, A., Wang, J.-X., {et~al.} 2018, Monthly Notices of the Royal
  Astronomical Society, 479, 5022

\bibitem[{Liu {et~al.}(2016)Liu, Merloni, Georgakakis, Menzel, Buchner, Nandra,
  Salvato, Shen, Brusa, \& Streblyanska}]{Liu2016}
Liu, Z., Merloni, A., Georgakakis, A., {et~al.} 2016, MNRAS, 459, 1602

\bibitem[{Loh(2008)}]{Loh2008}
Loh, J.~M. 2008, ApJ, 681, 726

\bibitem[{Lyu \& Rieke(2018)}]{Lyu2018}
Lyu, J. \& Rieke, G.~H. 2018, The Astrophysical Journal, 866, 92

\bibitem[{Ma{\l}ek {et~al.}(2018)Ma{\l}ek, Buat, Roehlly, Burgarella, Hurley,
  Shirley, Duncan, Efstathiou, Papadopoulos, Vaccari, Farrah, Marchetti, \&
  Oliver}]{Malek2018}
Ma{\l}ek, K., Buat, V., Roehlly, Y., {et~al.} 2018, Astronomy {\&}
  Astrophysics, 620, A50

\bibitem[{Malizia {et~al.}(2020)Malizia, Bassani, Stephen, Bazzano, \&
  Ubertini}]{Malizia2020}
Malizia, A., Bassani, L., Stephen, J.~B., Bazzano, A., \& Ubertini, P. 2020,
  Astronomy {\&} Astrophysics, 639, A5

\bibitem[{Masoura {et~al.}(2020)Masoura, Georgantopoulos, Mountrichas, Vignali,
  Koulouridis, Chiappetti, Fotopoulou, Paltani, \& Pierre}]{Masoura2020}
Masoura, V.~A., Georgantopoulos, I., Mountrichas, G., {et~al.} 2020, Astronomy
  {\&} Astrophysics, 638, A45

\bibitem[{Masoura {et~al.}(2021)Masoura, Mountrichas, Georgantopoulos, \&
  Plionis}]{Masoura2021}
Masoura, V.~A., Mountrichas, G., Georgantopoulos, I., \& Plionis, M. 2021,
  Astronomy {\&} Astrophysics, 646, A167

\bibitem[{Masoura {et~al.}(2018)Masoura, Mountrichas, Georgantopoulos, Ruiz,
  Magdis, \& Plionis}]{Masoura2018}
Masoura, V.~A., Mountrichas, G., Georgantopoulos, I., {et~al.} 2018, A\&A, 618,
  31

\bibitem[{Menzel {et~al.}(2016)}]{Menzel2016}
Menzel, M.-L. {et~al.} 2016, MNRAS, 457, 110

\bibitem[{Merloni {et~al.}(2014)Merloni, Bongiorno, Brusa, Iwasawa, Mainieri,
  Magnelli, Salvato, Berta, Cappelluti, Comastri, Fiore, Gilli, \&
  Koekemoer}]{Merloni2014}
Merloni, A., Bongiorno, A., Brusa, M., {et~al.} 2014, Monthly Notices of the
  Royal Astronomical Society, 437, 3550

\bibitem[{Mountrichas {et~al.}(2021{\natexlab{a}})Mountrichas, Buat, Yang,
  Boquien, Burgarella, \& Ciesla}]{Mountrichas2021}
Mountrichas, G., Buat, V., Yang, G., {et~al.} 2021{\natexlab{a}}, Astronomy
  {\&} Astrophysics, 646, A29

\bibitem[{Mountrichas {et~al.}(2021{\natexlab{b}})Mountrichas, Buat, Yang,
  Boquien, Burgarella, Ciesla, Malek, \& Shirley}]{Mountrichas2021b}
Mountrichas, G., Buat, V., Yang, G., {et~al.} 2021{\natexlab{b}}, Astronomy \&
  Astrophysics accepted [\eprint{2106.10678}]

\bibitem[{Mountrichas {et~al.}(2019)Mountrichas, Georgakakis, \&
  Georgantopoulos}]{Mountrichas2019}
Mountrichas, G., Georgakakis, A., \& Georgantopoulos, I. 2019, Monthly Notices
  of the Royal Astronomical Society, 483, 1374

\bibitem[{Mountrichas {et~al.}(2020)Mountrichas, Georgantopoulos, Ruiz, \&
  Kampylis}]{Mountrichas2020}
Mountrichas, G., Georgantopoulos, I., Ruiz, A., \& Kampylis, G. 2020, Monthly
  Notices of the Royal Astronomical Society, 491, 1727

\bibitem[{{Mountrichas} {et~al.}(2016)}]{Mountrichas2016}
{Mountrichas}, G. {et~al.} 2016, MNRAS, 457, 4195

\bibitem[{Mullaney {et~al.}(2015)Mullaney, Alexander, Aird, Bernhard, Daddi,
  Moro, Dickinson, Elbaz, Harrison, Juneau, Liu, Pannella, Rosario, Santini,
  Sargent, Schreiber, Simpson, \& Stanley}]{Mullaney2015}
Mullaney, J.~R., Alexander, D.~M., Aird, J., {et~al.} 2015, Monthly Notices of
  the Royal Astronomical Society: Letters, 453, L83

\bibitem[{Nenkova {et~al.}(2002)Nenkova, Ivezi{\'{c}}, \&
  Elitzur}]{Nenkova2002}
Nenkova, M., Ivezi{\'{c}}, {\v{Z}}., \& Elitzur, M. 2002, The Astrophysical
  Journal, 570, L9

\bibitem[{Netzer(2015)}]{Netzer2015}
Netzer, H. 2015, Annual Review of Astronomy and Astrophysics, 53, 365

\bibitem[{Ogawa {et~al.}(2021)Ogawa, Ueda, Tanimoto, \& Yamada}]{Ogawa2021}
Ogawa, S., Ueda, Y., Tanimoto, A., \& Yamada, S. 2021, The Astrophysical
  Journal, 906, 84

\bibitem[{{Oliver} {et~al.}(2012)}]{Oliver2012}
{Oliver}, S.~J. {et~al.} 2012, MNRAS, 424, 1614

\bibitem[{Park {et~al.}(2006)Park, Kashyap, Siemiginowska, van Dyk, Zezas,
  Heinke, \& Wargelin}]{Park2006}
Park, T., Kashyap, V.~L., Siemiginowska, A., {et~al.} 2006, The Astrophysical
  Journal, 652, 610

\bibitem[{{Pierre} {et~al.}(2016)}]{Pierre2016}
{Pierre}, M. {et~al.} 2016, A\&A, 592, 1

\bibitem[{Pouliasis {et~al.}(2020)Pouliasis, Mountrichas, Georgantopoulos,
  Ruiz, Yang, \& Bonanos}]{Pouliasis2020}
Pouliasis, E., Mountrichas, G., Georgantopoulos, I., {et~al.} 2020, Monthly
  Notices of the Royal Astronomical Society, 495, 1853

\bibitem[{Prevot {et~al.}(1984)Prevot, Lequeux, Maurice, Prevot, \&
  Rocca-Volmerange}]{Prevot1984}
Prevot, M., Lequeux, J., Maurice, E., Prevot, L., \& Rocca-Volmerange, B. 1984,
  A\&A, 132, 389

\bibitem[{Schartmann {et~al.}(2008)Schartmann, Meisenheimer, Camenzind, Wolf,
  Tristram, \& Henning}]{Schartmann2008}
Schartmann, M., Meisenheimer, K., Camenzind, M., {et~al.} 2008, Astronomy {\&}
  Astrophysics, 482, 67

\bibitem[{Schreiber {et~al.}(2015)}]{Schreiber2015}
Schreiber, C. {et~al.} 2015, A\&A, 575, 29

\bibitem[{{Smee} {et~al.}(2013)}]{Smee2013}
{Smee}, S. {et~al.} 2013, AJ, 146, 32

\bibitem[{{Somerville} {et~al.}(2008){Somerville}, {Hopkins}, J., {Robertson},
  \& L.}]{Somerville2008}
{Somerville}, R.~S., {Hopkins}, P.~F., J., C.~T., {Robertson}, B.~E., \& L., H.
  2008, MNRAS, 391, 481

\bibitem[{Stalevski {et~al.}(2012)Stalevski, Fritz, Baes, Nakos, \&
  Popovi{\'{c}}}]{Stalevski2012}
Stalevski, M., Fritz, J., Baes, M., Nakos, T., \& Popovi{\'{c}}, L.~{\v{C}}.
  2012, Monthly Notices of the Royal Astronomical Society, 420, 2756

\bibitem[{Stalevski {et~al.}(2016)Stalevski, Ricci, Ueda, Lira, Fritz, \&
  Baes}]{Stalevski2016}
Stalevski, M., Ricci, C., Ueda, Y., {et~al.} 2016, Monthly Notices of the Royal
  Astronomical Society, 458, 2288

\bibitem[{{Stern} {et~al.}(2012){Stern}, {Assef}, {Benford}, {Blain}, {Cutri},
  {Dey}, {Eisenhardt}, {Griffith}, {Jarrett}, {Lake}, {Masci}, {Petty},
  {Stanford}, {Tsai}, {Wright}, {Yan}, {Harrison}, \& {Madsen}}]{Stern2012}
{Stern}, D., {Assef}, R.~J., {Benford}, D.~J., {et~al.} 2012, ApJ, 753, 30

\bibitem[{Trouille {et~al.}(2009)Trouille, Barger, Cowie, Yang, \&
  Mushotzky}]{Trouille2009}
Trouille, L., Barger, A.~J., Cowie, L.~L., Yang, Y., \& Mushotzky, R.~F. 2009,
  The Astrophysical Journal, 703, 2160

\bibitem[{Urry \& Padovani(1995)}]{Urry1995}
Urry, C.~M. \& Padovani, P. 1995, Publications of the Astronomical Society of
  the Pacific, 107, 803

\bibitem[{{Werner} {et~al.}(2004){Werner}, {Roellig}, {Low}, {Rieke}, {Rieke},
  {Hoffmann}, {Young}, {Houck}, {Brandl}, {Fazio}, {Hora}, {Gehrz}, {Helou},
  {Soifer}, {Stauffer}, {Keene}, {Eisenhardt}, {Gallagher}, {Gautier}, {Irace},
  {Lawrence}, {Simmons}, {Van Cleve}, {Jura}, {Wright}, \&
  {Cruikshank}}]{Werner2004}
{Werner}, M.~W., {Roellig}, T.~L., {Low}, F.~J., {et~al.} 2004, ApJS, 154, 1

\bibitem[{{Wright} {et~al.}(2010){Wright}, {Eisenhardt}, {Mainzer}, {Ressler},
  {Cutri}, {Jarrett}, {Kirkpatrick}, {Padgett}, {McMillan}, {Skrutskie},
  {Stanford}, {Cohen}, {Walker}, {Mather}, {Leisawitz}, {Gautier}, {McLean},
  {Benford}, {Lonsdale}, {Blain}, {Mendez}, {Irace}, {Duval}, {Liu}, {Royer},
  {Heinrichsen}, {Howard}, {Shannon}, {Kendall}, {Walsh}, {Larsen}, {Cardon},
  {Schick}, {Schwalm}, {Abid}, {Fabinsky}, {Naes}, \& {Tsai}}]{Wright2010}
{Wright}, E.~L., {Eisenhardt}, P.~R.~M., {Mainzer}, A.~K., {et~al.} 2010, AJ,
  140, 1868

\bibitem[{{Yan} {et~al.}(2013)}]{Yan2013}
{Yan}, L. {et~al.} 2013, AJ, 145, 55

\bibitem[{Yang {et~al.}(2020)Yang, Boquien, Buat, Burgarella, Ciesla, Duras,
  Stalevski, Brandt, \& Papovich}]{Yang2020}
Yang, G., Boquien, M., Buat, V., {et~al.} 2020, Monthly Notices of the Royal
  Astronomical Society, 491, 740

\bibitem[{Yang {et~al.}(2016)Yang, Brandt, Luo, Xue, Bauer, Sun, Kim, Schulze,
  Zheng, Paolillo, Shemmer, Liu, Schneider, Vignali, Vito, \& Wang}]{Yang2016}
Yang, G., Brandt, W.~N., Luo, B., {et~al.} 2016, The Astrophysical Journal,
  831, 145

\bibitem[{Zou {et~al.}(2019)Zou, Yang, Brandt, \& Xue}]{Zou2019}
Zou, F., Yang, G., Brandt, W.~N., \& Xue, Y. 2019, The Astrophysical Journal,
  878, 11

\end{thebibliography}
\bibliographystyle{aa}

\clearpage

\appendix

\section{Sources with problematic photometry and unsecure X-CIGALE classification}
\label{sec_appendix}

\begin{figure}
\centering
\begin{subfigure}[b]{0.5\textwidth}
   \includegraphics[width=1\linewidth, height=7.2cm]{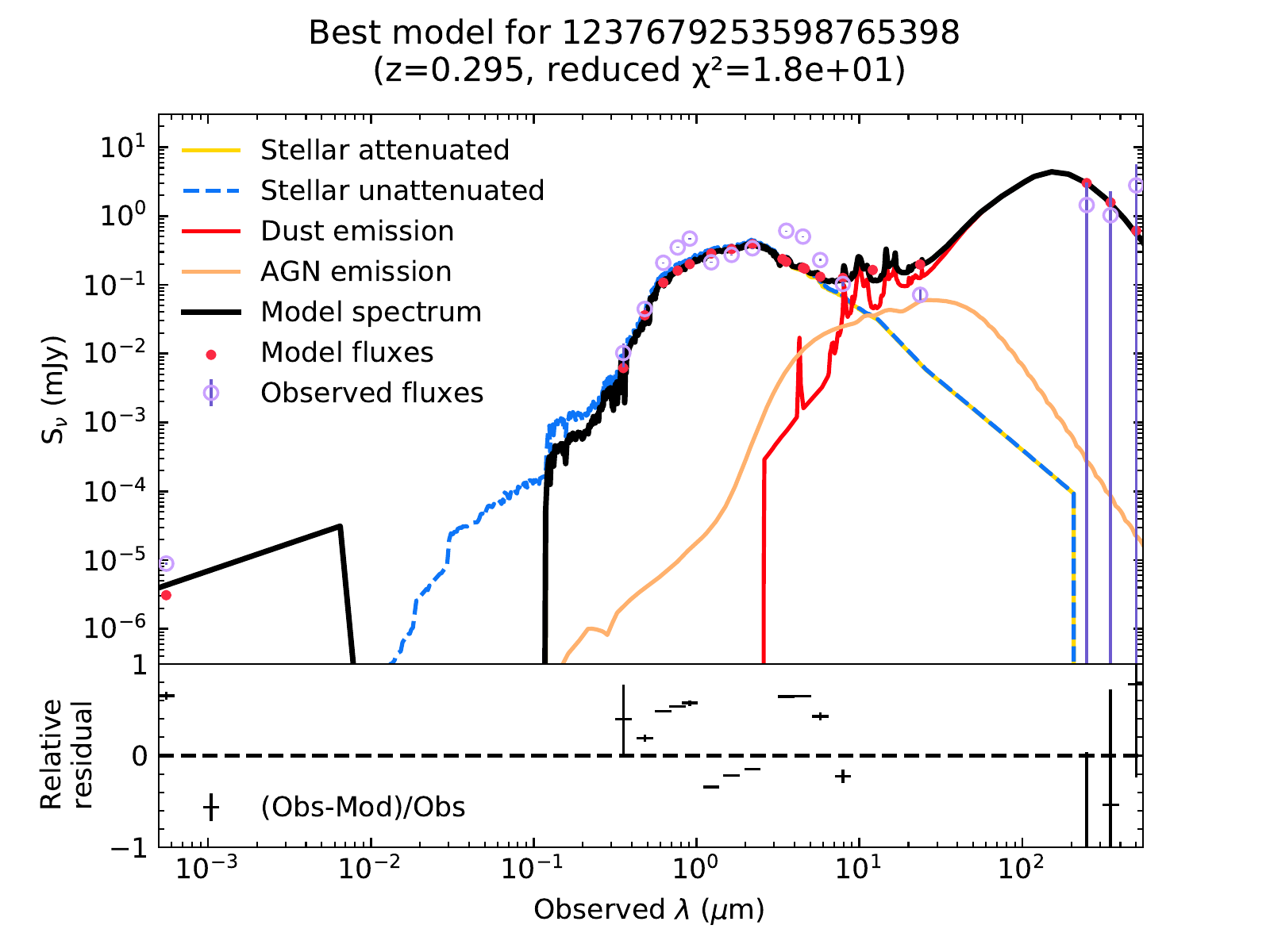}
   \label{} 
\end{subfigure}

\begin{subfigure}[b]{0.5\textwidth}
   \includegraphics[width=1\linewidth, height=7.2cm]{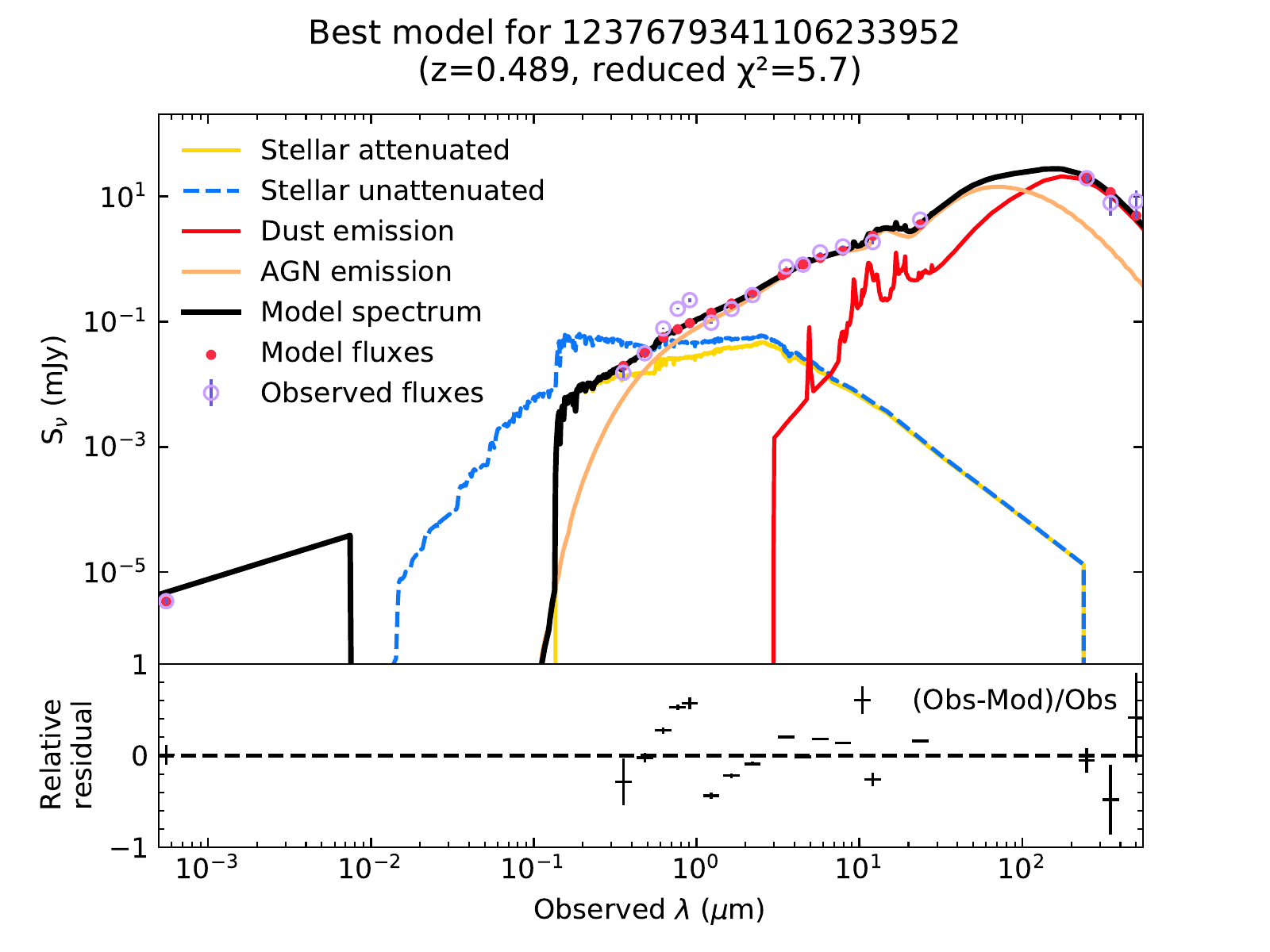}
   \label{}
\end{subfigure}

\caption{Examples of SEDs that have been excluded by our analysis, due to their problematic photometry that results in unreliable SED fits ($\chi ^2_{red}>5$).}
\label{fig_SEDs_problematic}
\end{figure}

From our analysis, we have excluded sources that have $\chi ^2_{red}>5$ (Section \ref{sec_unreliable}). Visual inspection of their SEDs shows that in these cases there are some problematic photometric bands that did now allow a reliable fit. Fig. \ref{fig_SEDs_problematic} presents two examples of these SEDs.

We also investigate further, the 44 sources that, although meet our selection requirements, they do not have secure classification from the SED fitting (Section \ref{sec_xcigale_classif}). Visual inspection of their SEDs reveals that X-CIGALE failed to provide a secure classification for at least one of the following reasons: one or more photometric data appear inconsistent, resulting to increased $\chi ^2_{red}$ values, but lower than the threshold we set to exclude sources. The majority of these sources has a low(er) AGN fraction compared to the rest of the population (mean frac$\rm_{AGN}=0.28$, compared to frac$\rm_{AGN}=0.43$ for those with secure classification). The UV/optical continuum is dominated by the stellar component, rendering hard to distinguish between a type 2 AGN and a type 1 AGN with increased polar dust (see below for the effect of polar dust on the classification). The latter is, in particular true, for these systems with low AGN fraction. Among the 44 AGN, 32 are spectroscopic type 1 and 12 are type 2. Sixteen out of the 32 have $i_{best}=30^{\circ}$, i.e., X-CIGALE identifies them as type 1, in agreement with their spectroscopic classification, but not securely since $i_{bayes}>40^{\circ}$. We fit again the remaining 16 sources, forcing them to be type 1 AGN ($i=30^{\circ}$). All 16 sources have $2.0<\Delta\rm BIC<4.0$. This suggests that there is no statistical difference in the SED fits, regardless of whether these sources are fitted as type 1 or type 2. The twelve sources that are type 2, but do not have secure classification from their SED fitting, present similar results. Specifically, 5/12 have $i_{best}=70^{\circ}$. For the remaining seven, we run X-CIGALE again and force them to be type 2 ($i=70^{\circ}$) and compare the fits from the two runs. $\Delta\rm BIC$ analysis shows that there is no strong preference in favour of either of the two fits. We conclude that the X-CIGALE classification for these 44 AGN is ambiguous.

\section{Classification of AGN based on X-CIGALE at z>1}

In the previous Sections, we restricted our analysis to those sources that lie at $\rm z<1$. This was due to the fact that there are no AGN classified as type 2 based on optical spectra, at higher redshift (Section \ref{sec_optspectra}). In this Section, we present the X-CIGALE classification of sources at $\rm z>1$ and compare it with spectroscopically classified type 1 AGN for which there is available information from the \cite{Menzel2016} catalogue.

There are 978 X-ray AGN that satisfy our photometric criteria, have a secure optical classification (Section \ref{sec_sample}) and  $\chi ^2_{red}<5$ from their SED fitting. 785 of them ($\sim 80\%$) have secure classification. This is similar to the fraction of sources that have secure classification at $\rm z<1$ ($\sim 85\%$, Section \ref{sec_xcigale_classif}). 566 AGN are classified by X-CIGALE as type 1 and 219 as type 2. Thus, X-CIGALE classification agrees with that from optical spectra for 72\% ($566/785$) of the sources. This number, is somewhat lower compared to the percentage of sources that are classified as type 1, by X-CIGALE, and are also spectroscopic type 1, at $\rm z<1$ ($160/188\approx 85\%$, Section \ref{sec_xcigale_classif}). 

We note that, at high redshifts, a larger fraction of sources is missing photometric data. Among the 240 AGN at $\rm z<1$ (Section \ref{sec_xcigale_classif}), 80\% have near-IR observations. However, at $\rm z>1$, 37\% of the sources lack near-IR data (48\% among the 219 that are classified as type 2, at $\rm z>1$, by X-CIGALE). Similarly, 75\% of the 240 AGN have been observed by {\it{Herschel}}. This fraction drops to 55\% at $\rm z>1$ and decreases further among the 219 AGN classified as type 2 (27\%). It is also worth pointing out that the 219 AGN that are spectroscopic type 1, but classified as type 2 by X-CIGALE, lie at a mean redshift of $\rm z=2.0$, i.e. higher than that of the 566 sources that are classified as type 1, both from optical spectroscopy and SED fitting ($\rm z=1.6$). 



We run X-CIGALE forcing these 219 sources to be classified as type 1, in accordance with their spectroscopic classification. We then compare the fits from this run with that when the classification is free. For that, we estimate the $\rm \Delta BIC$ parameter. With the exception of three sources that have $\rm \Delta BIC>6.0$, $\rm \Delta BIC$ varies from two to six, which indicates that the two fits do not differ statistically. Thus, the 219 AGN could also be classified as type 1, in agreement with their spectroscopic classification. 

From the 219 sources, we select 20 of them that satisfy the photometric criteria presented in Section \ref{sec_sample} and additionally have near-IR and {\it{Herschel}} observations. Similarly to the results we found at $\rm z<1$, for the 28 spectroscopic type 1, X-ray AGN classified as type 2 by X-CIGALE, the majority of the sources (12/20) present increased polar dust ($\rm E_{(B-V), bayes}>0.15$). The mean $\rm E_{(B-V), bayes}$ increases, from $\rm E_{(B-V), bayes}=0.12$ to $\rm E_{(B-V), bayes}=0.19$, when we force these AGN to be fitted as type 1 . Two sources have N$_H>21.5$\,cm$^{-2}$. 

We conclude that, X-CIGALE can classify type 1 AGN at $\rm z>1$ with the same efficiency as at lower redshifts, under the condition that there is sufficient photometric coverage at high redshifts.

\end{document}